\documentclass{aa}  

\usepackage{natbib}
\bibpunct{(}{)}{;}{a}{}{,} 
\usepackage{graphicx}
\usepackage{txfonts}
\usepackage{twoopt}
\usepackage[breaklinks=true,colorlinks=true,allcolors=blue]{hyperref}
\usepackage{pifont}
\usepackage{array}
\usepackage{adjustbox}
\usepackage{enumitem}
\usepackage{xcolor}
\usepackage{placeins}

\newcommand\kms{km~s$^{-1}$}
\newcommand\msun{$M_\odot$}
\newcommand\mhi{$M_{\rm HI}$}

\newcommand\vcirc{$V_{\rm circ}$}
\def\be{\begin{equation}}
\def\ee{\end{equation}}
\def\a40{$\alpha$.40}
\def\arcmin{$^{\prime}$}
\def\arcsec{$^{\prime\prime}$}
\def\dg{$^{\circ}$}

\def\hi{H\,{\sc i}}
\def\mstar{$M_{*}$}

\def\w50{$W_{50}$}
\def\wtw{$W_{20}$}

\def\PAkin{$PA_{\rm kin}$}
\def\PAop{$PA_{\rm op}$}
\def\vrot{$V_{\rm rot}$}

\def\darkgal{J0139+4328}

\usepackage{natbib,twoopt}
\usepackage[breaklinks=true]{hyperref} 
\bibpunct{(}{)}{;}{a}{}{,} 
\makeatletter
\newcommandtwoopt{\citeads}[3][][]{\href{http://adsabs.harvard.edu/abs/#3}%
{\def\hyper@linkstart##1##2{}%
\let\hyper@linkend\@empty\citealp[#1][#2]{#3}}}
\newcommandtwoopt{\citepads}[3][][]{\href{http://adsabs.harvard.edu/abs/#3}%
{\def\hyper@linkstart##1##2{}%
\let\hyper@linkend\@empty\citep[#1][#2]{#3}}}
\newcommandtwoopt{\citetads}[3][][]{\href{http://adsabs.harvard.edu/abs/#3}%
{\def\hyper@linkstart##1##2{}%
\let\hyper@linkend\@empty\citet[#1][#2]{#3}}}
\newcommandtwoopt{\citeyearads}[3][][]%
{\href{http://adsabs.harvard.edu/abs/#3}
{\def\hyper@linkstart##1##2{}%
\let\hyper@linkend\@empty\citeyear[#1][#2]{#3}}}
\makeatother

\begin{document}

\title{Not so-dark: High resolution \hi\ imaging of J0139+4328 and identification of an optical counterpart}

\author{Barbara Šiljeg\inst{1,2}
        \and
        Elizabeth A.\ K.\ Adams\inst{1,2}
        \and 
        Tom A.\ Oosterloo\inst{1,2}
        \and
        Filippo Fraternali\inst{2}
        \and 
        Kelley M. Hess\inst{3,1}
        \and
        Jin-Long Xu\inst{4,5,6}
        \and
        Ming Zhu\inst{4,5,6}
        }
\institute{
    ASTRON, Netherlands Institute for Radio Astronomy, Oude Hoogeveensedijk 4, 7991 PD Dwingeloo, The Netherlands
    \and
    Kapteyn Astronomical Institute, University of Groningen Postbus 800, 9700 AV Groningen, the Netherlands
    \and
    Department of Space, Earth and Environment, Chalmers University of Technology, Onsala Space Observatory, 43992 Onsala, Sweden
    \and
    National Astronomical Observatories, Chinese Academy of Sciences, Beijing 100101, People's Republic of China
    \and
    Guizhou Radio Astronomical Observatory, Guizhou University, Guiyang 550000, People's Republic of China
    \and 
    CAS Key Laboratory of FAST, National Astronomical Observatories, Chinese Academy of Sciences, Beijing 100101, People's Republic of China
}


\abstract{
Dark galaxies -- systems rich in neutral hydrogen (\hi) gas but with no stars -- are a common prediction of numerous theoretical models and cosmological simulations. However, the unequivocal identification of such sources in current \hi\ surveys has proven challenging. 
In this work, we present interferometric follow-up observations with the VLA of a former dark galaxy candidate \darkgal, originally detected with the single-dish FAST telescope. The improved spatial resolution of the VLA data allow us to identify a faint optical counterpart and characterize the galaxy. Located at a distance of about 31 Mpc, \darkgal\ has a stellar mass of $3\times 10^6$\msun\ and a relatively high gas richness of \mhi/\mstar\ = 18. Despite its high ratio, the galaxy is consistent, within the scatter, with the stellar-to-\hi\ mass relation of \hi-selected samples in the literature and with the baryonic Tully–Fisher relation (BTFR), although its kinematic measurement is subject to large uncertainties.
This case highlights the potential of modern high-sensitivity \hi\ surveys for detecting low surface brightness, gas-rich galaxies, but underscores the need for careful interpretation of low-resolution \hi\ data, with potentially large centroid errors, and for sufficiently deep optical imaging to ensure robust identification.
}

\date{}
\titlerunning{\darkgal: VLA follow-up and optical counterpart}
\authorrunning{B. Šiljeg et al.}
\maketitle


\section{Introduction}\label{p3:sec:intro}

The search for dark galaxies, that is, galaxies containing gas but no
stars, has been of interest for many decades, dating back to early surveys of the neutral hydrogen (\hi) content of galaxies \citepads[e.g.][]{2002ApJS..142..161P,2005MNRAS.361...34D,2006MNRAS.368.1479D}.
Theoretically, these systems are thought to be dark matter halos in which neutral gas could not reach the critical density required to trigger star formation. This is attributed to a combination of poor cooling efficiency in low-metallicity environments and the suppression of gas collapse due to the cosmic ultraviolet (UV) background radiation \citepads[e.g.][]{1997ApJ...487...61K}. Some analytic models \citepads[e.g.][]{2006MNRAS.368.1479D,2020MNRAS.498.4887B} and cosmological simulations \citepads[e.g.][]{2017MNRAS.465.3913B,2024ApJ...962..129L} predict the existence of such systems and provide further insight into their evolution. In particular, \citetads{2024ApJ...962..129L} found that dark galaxies in the Illustris TNG50 simulation \citepads{2014Natur.509..177V,2015A&C....13...12N} tend to reside in dark matter halos with slightly higher spin parameters, have larger gas and dark matter sizes compared to luminous systems of similar mass, and are found in lower-density regions where reduced number of interactions allow them to persist without merging with luminous systems. Finding such systems in observations serves as a direct test of current theoretical models when compared to predictions from simulations.

While the existence of dark galaxies is theoretically anticipated, there are very few convincing dark galaxy candidates to date.
Numerous \hi-rich but optically faint or undetected sources have been identified in untargeted \hi\ surveys such as the Arecibo Legacy Fast ALFA (ALFALFA; \citeads{2005AJ....130.2598G}) and the FAST all sky \hi\ survey (FASHI, \citeads{2024SCPMA..6719511Z}). 
However, most previous candidates were later revealed to more likely be remnants of tidal interactions or gas stripping (e.g. VIRGOHI21, \citeads{2005MNRAS.363L..21B,2017MNRAS.467.3648T}; SECCO 1, \citeads{2015A&A...575A.126B,2015ApJ...800L..15B,2018MNRAS.476.4565B}).
Some most promising current candidates include HI 1225+01 \citepads{1989ApJ...346L...5G,1995AJ....109.2415C,2012AJ....144..159M}, 
AGESVC1 282 \citepads{2012MNRAS.423..787T,2013MNRAS.428..459T,2020A&A...642L..10B}, 
Cloud 9 \citepads{2023ApJ...952..130Z,2023ApJ...956....1B,2024ApJ...973...61B}, 
and AC G185.0-11.5 \citepads{2025SciA...11S4057L}.
However, all of the above candidates are close to another galaxy or belong to a cluster, complicating their interpretations. The lack of truly isolated candidates questions the existence of these systems and/or their detectability with current \hi\ surveys. Recent untargeted \hi\ surveys such as the APERture Tile In Focus (Apertif;  \citeads{2022A&A...658A.146V,2022A&A...667A..38A}) and the Widefield ASKAP L-band Legacy All-sky Blind surveY (WALLABY; \citeads{2020Ap&SS.365..118K}) are continuing to expand the search for such systems, with some already proposed candidates (with WALLABY; \citeads{2025arXiv250504299O}) which, however, are yet to be confirmed.

Despite the scarcity of unequivocal dark galaxies, deep \hi\ surveys have uncovered a variety of systems that are extremely gas-rich and under luminous. These include low-mass galaxies with unusually high gas richness (e.g. Coma P, \citeads{2015ApJ...801...96J}; AGC 229101, \citeads{2021AJ....162..274L}), and ultra-diffuse galaxies (UDGs), galaxies of low surface brightnesses ($\mu$ > 24 mag arcsec$^{-2}$) but large physical sizes ($R_e$ > 1.5 kpc; e.g. \citeads{2017ApJ...842..133L,2019ApJ...883L..33M,2024A&A...692A.217S}). 
Together, these objects define a broader regime of interest in galaxy evolution—probing the physical conditions under which dark matter halos can accumulate baryons, and the circumstances in which those baryons fail to form stars efficiently. The study of such galaxies offers a promising path to refining models of gas accretion, cooling, and feedback in low-mass systems \citepads[e.g.][]{2015ARA&A..53...51S}.

In this work, we report follow-up observations of a galaxy that was initially discovered with FAST and reported by \citetads{2023ApJ...944L..40X} (hereafter X23) as a dark galaxy candidate. Using interferometric \hi\ observations from the Karl G. Jansky Very Large Array (VLA), we were able to refine the position of the \hi\ emission and identify the faint optical counterpart in the Pan-STARRS1 (PS1) data \citepads{2016arXiv161205560C}. The system remains under luminous, with a log (\mhi /$L_{V}$) of 0.94, comparable to ultra diffuse galaxies \citepads{2017ApJ...842..133L,2019ApJ...883L..33M,2019MNRAS.490..566J,2021AJ....162..274L}, but at lower masses. When compared with larger samples of \hi-selected galaxies, we find that \darkgal\ is not an outlier from the \mhi-\mstar\ scaling relation, although it lies in a relatively unexplored regime of dwarf galaxies with low stellar mass and high gas richness. This case highlights the challenges of associating optical counterparts to \hi\ detections from low resolution single-dish surveys, particularly at larger distances where the optical counterpart is expected to be faint in existing, large-scale optical surveys.

In Sect. \ref{p3:sec:data}, we describe the VLA observations and ground-based imaging data. In Sect. \ref{p3:sec:properties}, we derive and report the global \hi\ and optical properties of the galaxy, and compare with previous \hi\ detections of the galaxy with FAST. We report on the environment of \darkgal\ and place it into context of other \hi\ detected galaxies in Sect. \ref{p3:sec:discussion}. Finally, in Sect. \ref{p3:sec:conclusions_and_outlook}, we provide our conclusions and considerations for future studies of dark galaxies.

\section{Data}\label{p3:sec:data}

\subsection{Resolved \hi\ observations}

Our target \darkgal\ was observed with the Karl G. Jansky Very Large Array (VLA) in D-configuration using exploratory time under program 23A-424. The observations were done in four blocks of 3 hours each on December 20, 21, 22 and 27, 2023, with 8.8 hours in total on \darkgal. In addition, observations using the C-array configuration were done under project 25A-251 on July 29 and 30 and August 2 an 10 with a total on target observing time of 11.7 hr.
In all observations, the flux- and bandpass calibrator 3C48 was observed for 15 minutes at the start of each observing block. \darkgal\ was observed as a set of 15-minutes scans interleaved with 5-minute scans on the secondary calibrator J0136+4751. The spectral setup consisted of an 8 MHz bandwidth with 1024 channels giving a channel width of 7.81 kHz corresponding to 1.65 \kms\ (radio convention).
The data were flagged for minor radio frequency interference and standard cross-calibrations using the primary and the phase calibrator were applied to the target data. No self-calibration was performed. The data from both D- and C-array configurations were combined and imaged with a robust weighting of --0.5 and a channel width of 1.65 \kms. Hanning smoothing was applied to the data cube resulting in a velocity resolution of 3.3 \kms. The continuum was subtracted by fitting a straight line through the line-free channels. The noise level of the spectral line cube is 0.64 mJy beam$^{-1}$ and the beam size is 21\farcs8$\times$ 15\farcs5 (position angle 48\fdg1). This gives a 3-$\sigma$ column density sensitivity of $3.8 \times 10^{19}$ cm$^{-2}$ for a line width of 10 \kms. The line cube  was initially cleaned without a mask down to 5 times the noise level. The result of this was smoothed to twice the spatial resolution and with a 5-channel wide Hanning function in velocity. A mask was created from this smoothed cube using a clip level of twice the noise level of this lower-resolution cube. The original line cube was then cleaned using this mask down to a level of half the noise level.

After determining the systemic velocity of the source (Sect. \ref{p3:sec:HI_properties}), we transfer the cube to the rest frame of the source using:
\begin{equation}
    V_{\text{source frame}} = \frac{\nu_{\rm sys} - \nu_{\rm obs}}{\nu_{\rm sys}} \cdot c
\end{equation}
where $\nu_{\rm sys}$ is the systemic frequency of the source, $\nu_{\rm obs}$ the observed frequency of the spectral line emission, and $c$ the speed of light. The channel width and spectral resolution in this rest frame become $1.66$ \kms\ and 3.33 \kms, respectively. All \hi\ properties of \darkgal\ and all plots showing the \hi\ emission are given in the rest frame of the source.

\subsection{Pan-STARRS1}
We used PS1 data \citepads{2016arXiv161205560C} to characterize the stellar counterpart as this is the only optical photometric survey with publicly available data in this region. PS1 is a broadband photometric survey conducted using the 1.8-meter telescope located at the Haleakalā Observatory in Hawaii. We use the PS1 cutout service to obtain 
3.3\arcmin$\times$3.3\arcmin\ images in $g$-, $r$-, and $i$-bands centered at the \hi\ centroid position (see Sect. \ref{p3:sec:HI_properties}). The median seeings in $g$-, $r$-, and $i$-bands are estimated by Gaussian fits to foreground stars and are equal to 1.27\arcsec, 1.11\arcsec, and 1.03\arcsec, respectively.


\section{Properties of \darkgal}
\label{p3:sec:properties}

\begin{table}[]
    \centering
    \caption{{Properties of \darkgal}}
    \begingroup

    \renewcommand{\arraystretch}{1.3}
    \begin{tabular}{ll}
    \hline \hline
    Property & Value \\
    \hline
    \multicolumn{2}{c}{\hi} \\
    \hline
    centroid    & 01:39:29.7 +43:28:30 \\
    $V_{\rm sys}^{\rm hel,op}$ & ($2483.5 \pm 0.5$) \kms\\
    $D$ & ($31.03 \pm 5.6$) Mpc\\
    $f_c$ & 150 pc arcsec$^{-1}$ \\
    $S_{\text{\hi}}$ & ($259 \pm 26$) mJy \kms\\
    \mhi & $(5.9 \pm 2.2) \times 10^7$ \msun \\
    \w50 & ($28.5 \pm 1.2$) \kms\\
    \wtw & ($43.4 \pm 1.9$) \kms\\
    \PAkin & $\sim$150\dg \\

    \hline
    \multicolumn{2}{c}{Optical} \\
    \hline
    centroid    &  01:39:29.5 +43:28:30\\
    $i$ & $(56 \pm 7)$\dg\\
    
    \PAop & $(128 \pm 2)$\dg\\
    
    $m_g$ & $20.55_{-0.26}^{+0.35}$\\

    $m_r$ & $20.32_{-0.18}^{+0.21}$\\

    $m_i$ & $20.31_{-0.27}^{+0.38}$\\

    \mstar &  $3.2_{-2.0}^{+6.4} \times 10^6$ \msun\\
    
    \hline
    \end{tabular}
    \endgroup
    \tablefoot{Systemic velocity ($V_{\rm sys}^{\rm hel,op}$) is given in heliocentric rest frame using optical convention. All \hi\ properties are reported in the rest frame of the source. $f_c$ is the conversion factor from arcsec to pc. \PAkin\ is defined as the angle of the receding side measured counterclockwise from the north direction.}
    \label{p3:tab:props}
\end{table}

\begin{figure*}
    \centering
    \includegraphics[width=0.45\linewidth]{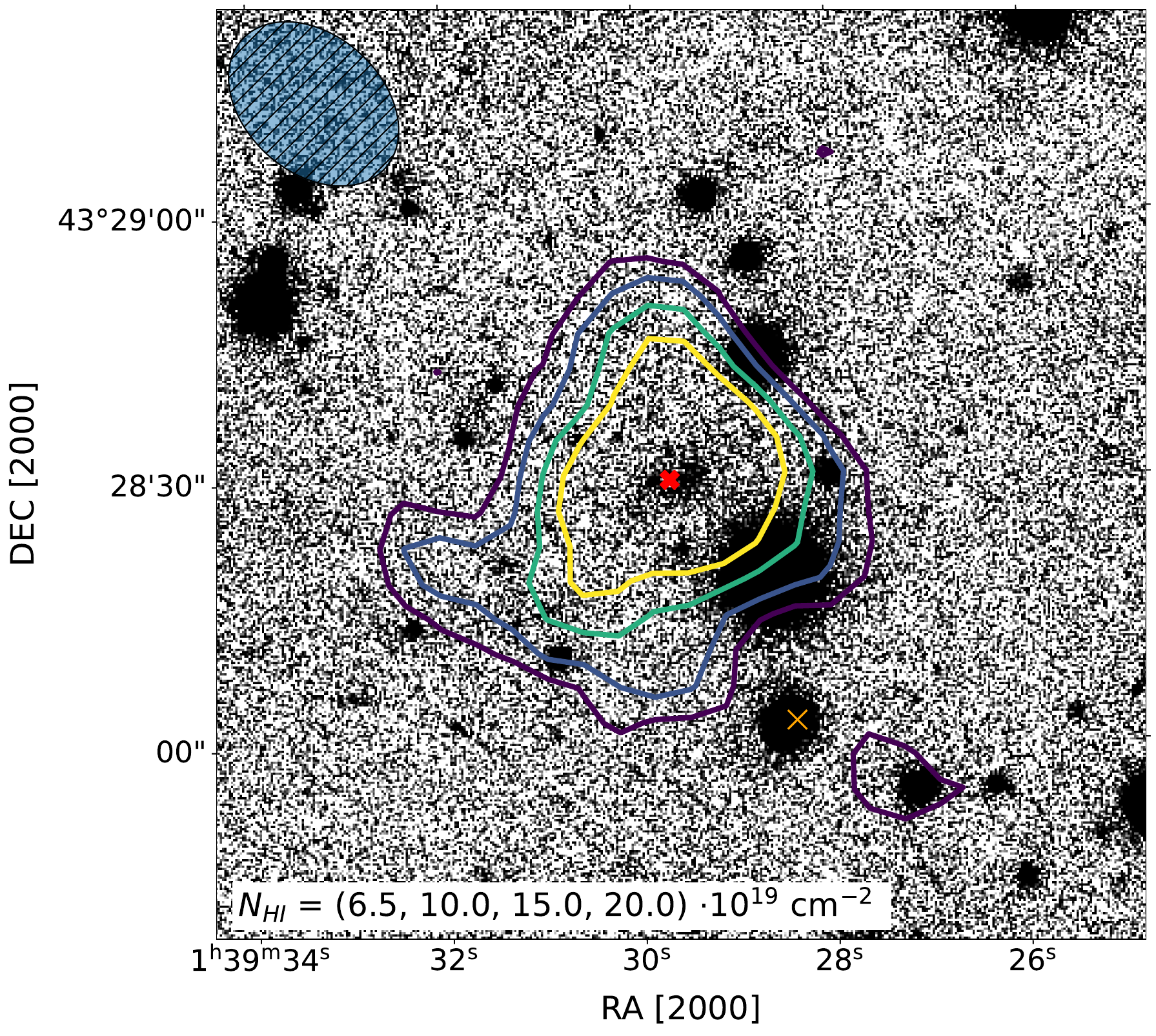}
    \centering
    \includegraphics[width=0.45\linewidth]{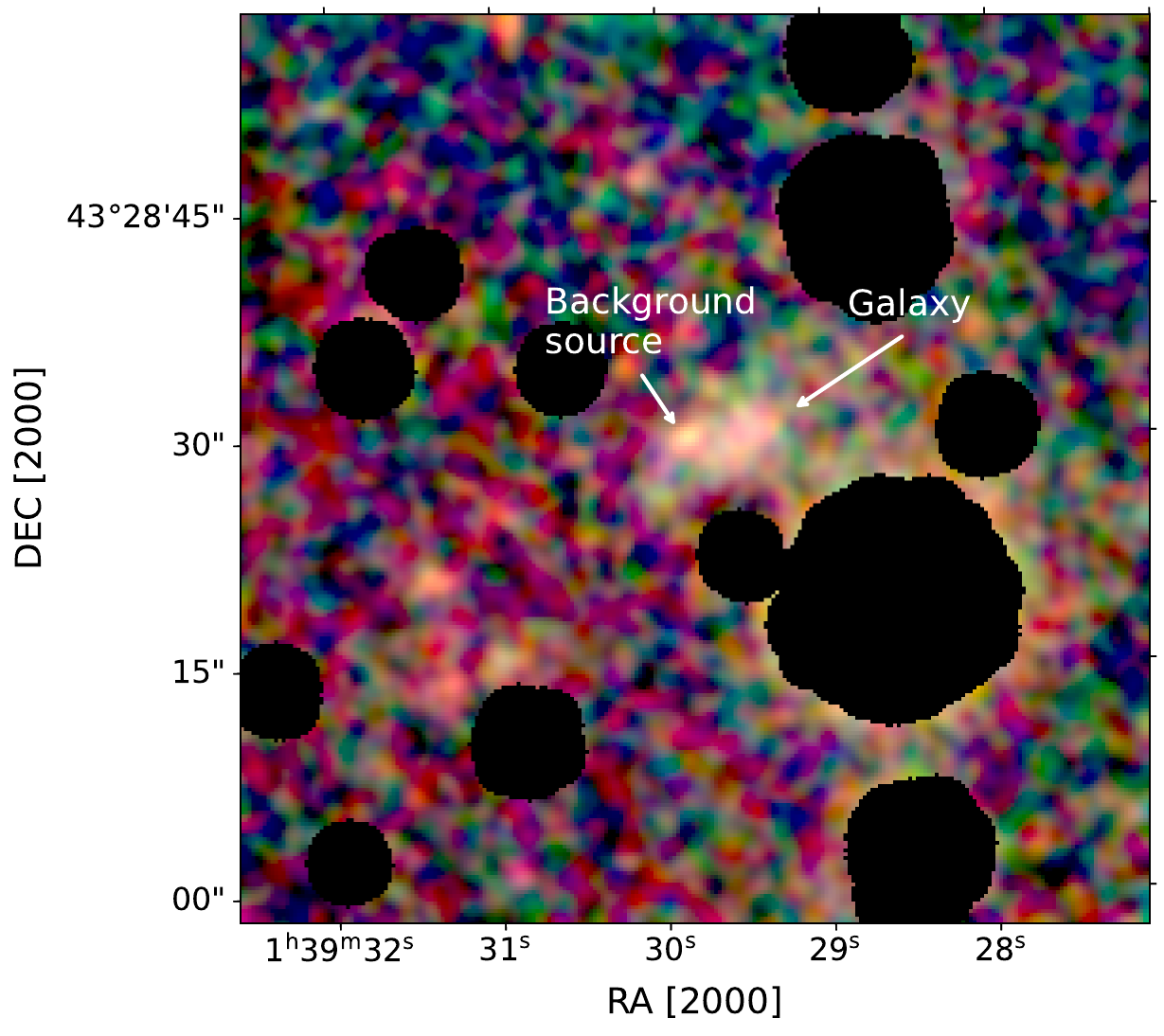}
    \caption{Identification of the optical counterpart. \textit{Left:} \hi\ contours on optical image obtained by stacking PS1 $g$, $r$, and $i$-band images. The \hi\ contours are [6.5, 10, 15, 20] $\times$ $10^{19}$ atoms cm$^{-2}$, where the lowest contour is $\sim$2-$\sigma$. The thick red cross marks the VLA centroid, while the thin orange cross marks the FAST centroid from X23. \textit{Right:} Color image of the galaxy from $g$ (blue), $r$ (green), and $i$-bands (red), smoothed by a Gaussian kernel of 2 pixels. Masked areas are in black.}
    \label{p3:fig:hiopt}
\end{figure*}

\begin{figure}
    \centering
    \includegraphics[width=0.9\linewidth]{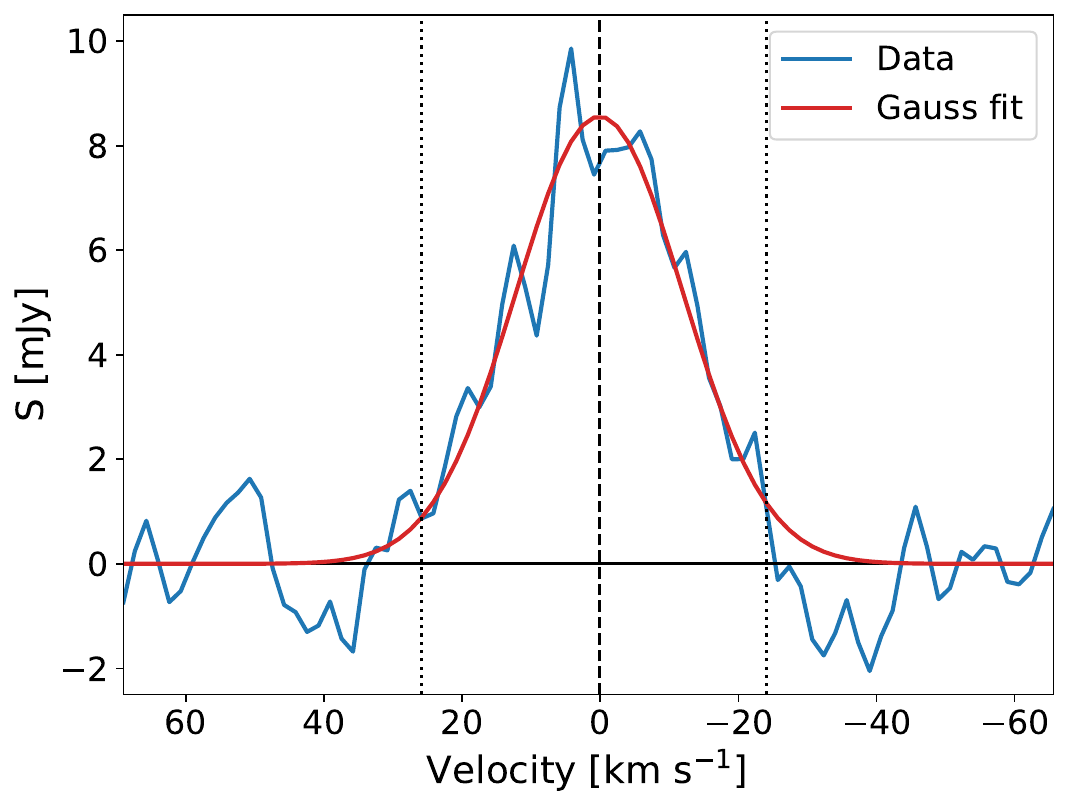}
    \caption{VLA global \hi\ profile of \darkgal\ in the rest frame of the source. Dotted lines indicate the extent used for creating the moment zero map and the dashed line denotes the determined systemic velocity.}
    \label{p3:fig:spec}
\end{figure}

\begin{figure}
    \centering
    \includegraphics[width=1\linewidth]{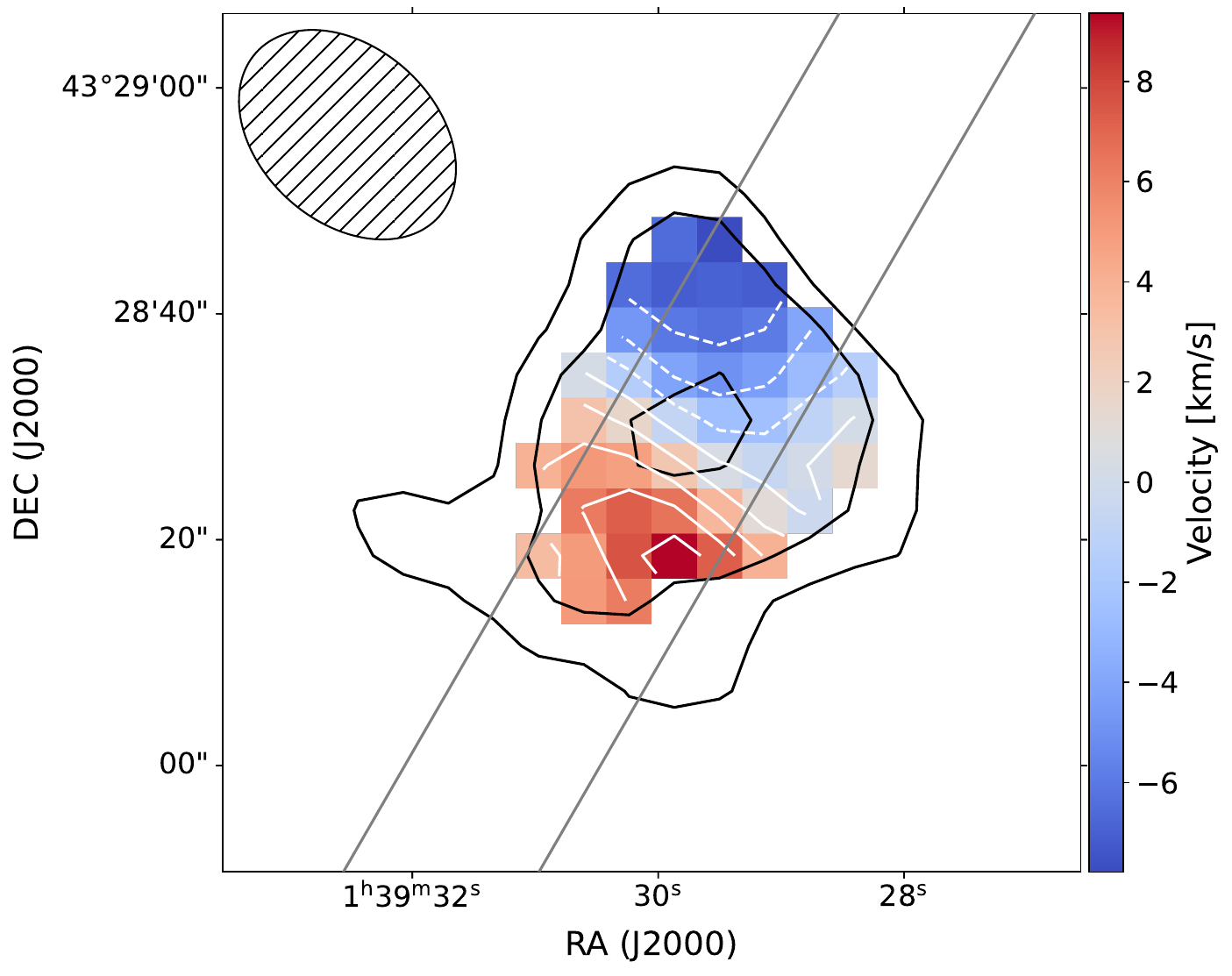}
    \caption{Moment one map with total intensity \hi\ contours at [3, 5, 10]-$\sigma$ overlaid in black. White contours are iso-velocity contours separated by 2 \kms. Full gray lines denote the area from which the position-velocity slice in Fig. \ref{p3:fig:pvslice} was extracted.}
    \label{p3:fig:mom1}
\end{figure}

\begin{figure}
    \centering
    \includegraphics[width=1\linewidth]{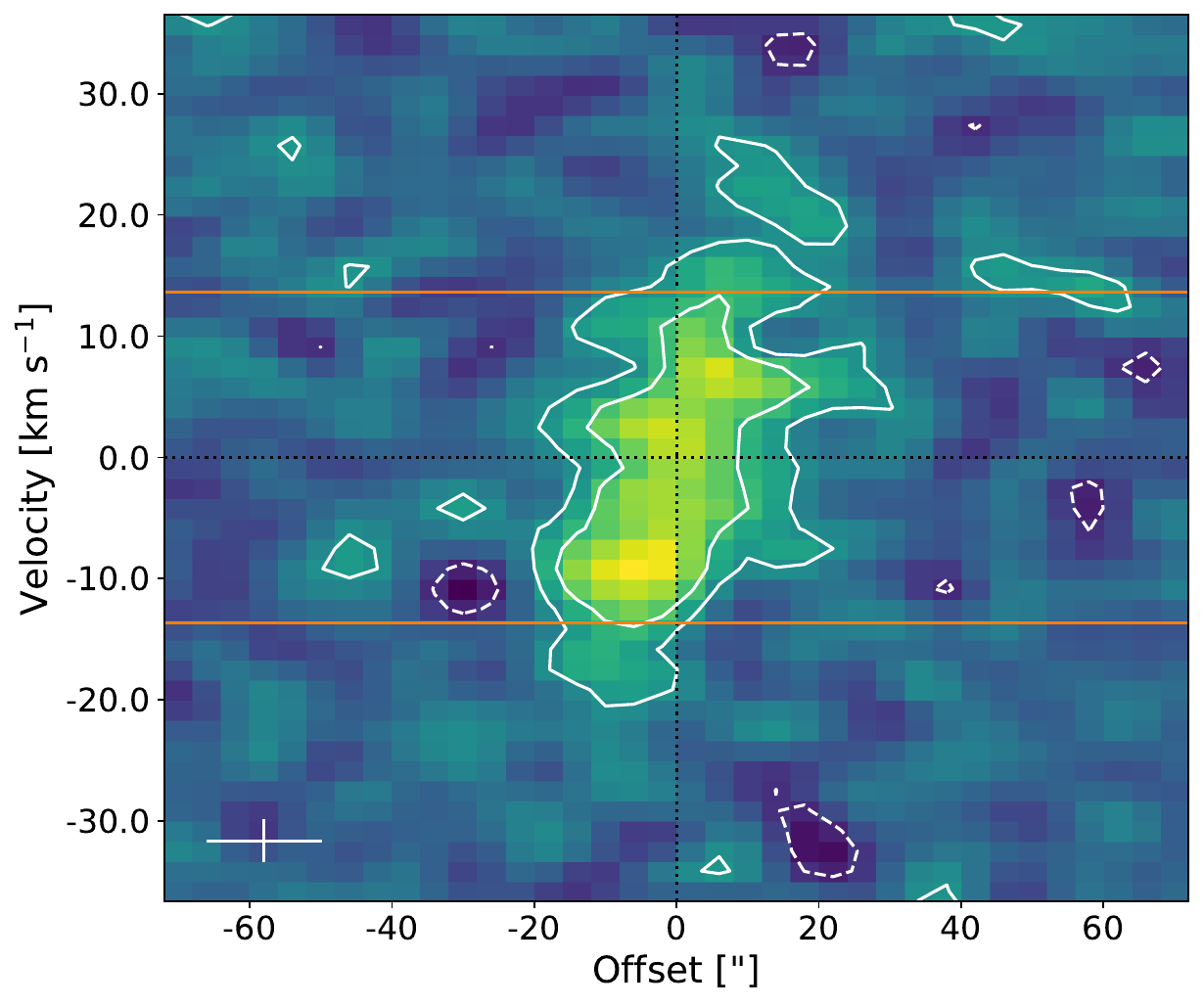}
    \caption{Position-velocity slice at the $PA = 150$\dg. Data contours are shown in white at [2, 4]-$\sigma$ with full (dashed) lines denoting positive (negative) emission. Orange horizontal lines denote the \w50\ spread. In the lower left corner, we denote the beam size (horizontal line) and velocity resolution (vertical line).}
    \label{p3:fig:pvslice}
\end{figure}

\begin{figure*}
    \centering
    \includegraphics[width=0.75\linewidth]{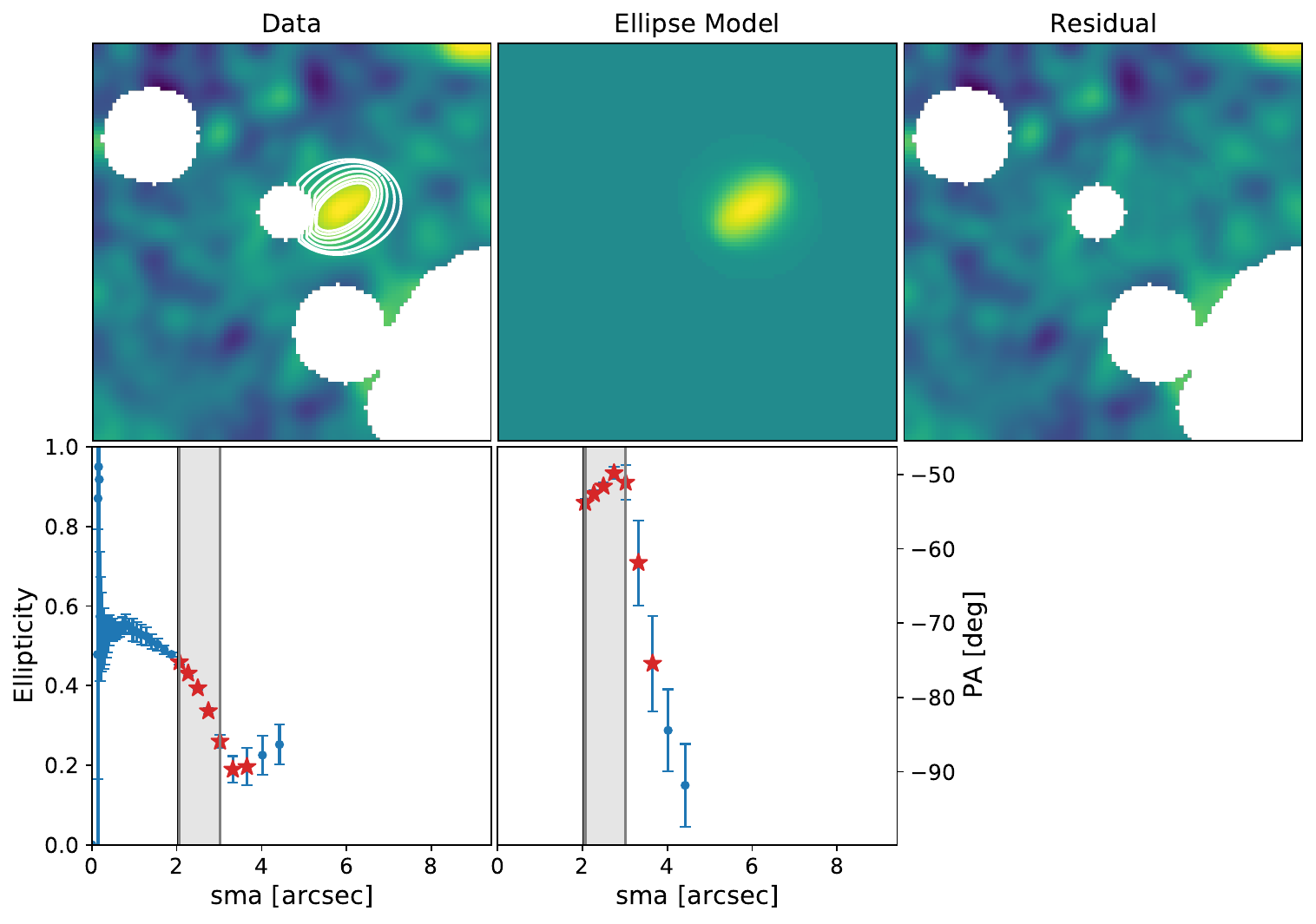}
    \caption{Isophotal fitting.
    \textit{Upper panels:} The left panel displays the smoothed image with overlaid ellipses whose corresponding parameters are indicated by red stars in the lower panels. Masked regions appear in white. The central panel presents the model reconstructed from all fitted ellipses whose parameters are shown as blue circles in the bottom panels, while the right panel shows the residual obtained by subtracting the model from the data image.
    \textit{Lower panels:} Ellipticity and position angle as functions of the semi-major axis length from the second iteration of the fitting procedure. Blue points represent all fitted ellipses from this run, and red stars correspond to those displayed in the upper left panel. The vertical black line indicates the radius equivalent to the PSF’s FWHM after smoothing, and the shaded gray area marks the region used to determine the overall geometry (by taking the median within this range).}
    \label{p3:fig:isophot_fit}
\end{figure*}

\begin{figure}
    \centering
    \includegraphics[width=0.9\linewidth]{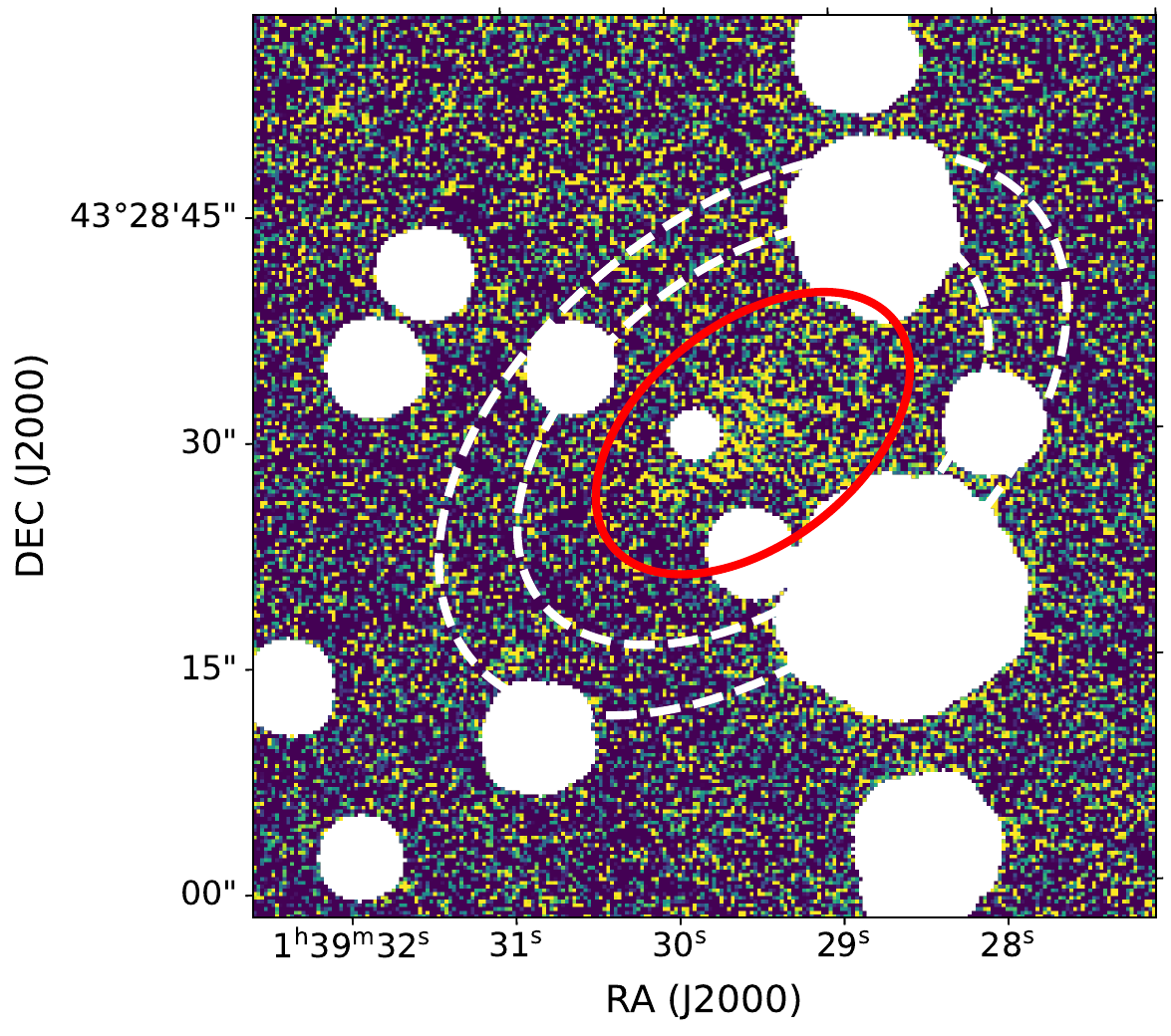}
    \caption{Aperture photometry on the masked image in the $r$-band. The red ellipse corresponds to the 12\arcsec\ (along the semi-major axis) aperture in which the magnitude was measured, while white dashed ellipses denote the annulus within which the background level was estimated at semi-major axes of 18\arcsec\ - 24\arcsec.}
    \label{p3:fig:photometry}
\end{figure}

In this section, we derive the \hi\ and optical properties of \darkgal. All derived quantities can be found in Table \ref{p3:tab:props}, while the methods are reported in the following.

\subsection{Distance}
\label{p3:sec:distance}
We derive the velocity-based distance to \darkgal\ using the calculators\footnote{\url{https://edd.ifa.hawaii.edu/}} provided by the Extragalactic Distance Database \citepads{2020AJ....159...67K}. We use the Numerical Action Method (NAM) model from \citetads{2017ApJ...850..207S} which is more precise for small distances (model extends to $\sim$ 38 Mpc). \darkgal\ has a heliocentric optical recessional velocity of 2484 \kms, corresponding to a velocity of 2651 \kms\ in the
Galactic standard of rest frame (eq. 4 from \citeads{2020AJ....159...67K}). The NAM model gives a distance of (31.03 $\pm$ 5.4) Mpc, where the error is estimated using the prescription from \citetads{2025A&A...696A.185H}.

\subsection{\hi\ properties}
\label{p3:sec:HI_properties}

We construct the moment zero map by collapsing all channels with visually identified emission (in the heliocentric rest frame and optical convention this would correspond to 2459.1 -- 2509.4 \kms; channel maps are shown in Appendix \ref{c3:app:chan_maps}). Fig. \ref{p3:fig:hiopt} shows the moment zero map overlaid on an optical image, with the peak emission clearly situated on top of an optical counterpart. We convert to column densities ($N_{HI}$) using:
\begin{equation}
    \frac{N_{HI}}{\text{cm}^{-2}} = 1.1\times 10^{24}\, \frac{S_{HI}}{\text{Jy \kms}}\, \frac{\text{arcsec}^2}{\theta_{\text{min}} \times \theta_{\text{maj}}},
\end{equation}
where $S_{HI}$ is the total \hi\ flux, and $\theta_{\text{min}}$ and $\theta_{\text{maj}}$ are the FWHMs of the minor and major axes of the beam, respectively. The \hi\ centroid is determined by fitting a 2D Gaussian to the innermost 24\arcsec$\times$24\arcsec\ region of the galaxy. We obtain the position of 01h39m29.7s +43d28m30s, offset by 31\arcsec\ from the X23 position obtained with FAST. The moment zero map was then smoothed to 30\arcsec\ resolution and clipped at 3-$\sigma$ to define a mask for the original resolution \hi\ cube, then applied to each channel. This procedure includes faint emission into the mask and has been shown to robustly recover the total \hi\ flux of a marginally resolved \hi\ source \citepads{2013A&A...555L...7O, 2018A&A...612A..26A}. Fig. \ref{p3:fig:spec} shows the global spectrum within this mask. We fit a Gaussian function to derive the \hi\ properties of the source (Table \ref{p3:tab:props}). Systemic velocity in the heliocentric rest frame and optical convention corresponds to 2483.5 \kms. All other \hi\ properties are derived in the rest frame of the source, and correspond to: velocity width of the global profile at 50\% of the peak value (\w50) of 28.5 \kms\ and total flux of 259 mJy \kms\footnote{This value is obtained by integrating the fitted Gaussian function and is perfectly consistent with the value obtained by integrating the data between -25.7 and 34.1 \kms\ where the spectrum is above zero.}. At the distance of 31.03 Mpc (Sect. \ref{p3:sec:distance}) the derived \hi\ mass of \darkgal\ is $5.9 \times 10^7$ \msun. Its peak column density, at a resolution of 16\arcsec$\times$22\arcsec\ is 3.5 $\times 10^{20}$ atoms cm$^{-2}$.

Fig. \ref{p3:fig:mom1} shows the moment one map, created over the same channel range as the moment zero map, with a mask applied based on a 5$\sigma$ (1.6$\times 10^{20}$ cm$^{-2}$) threshold in the moment zero map, in order to retain only high-significance emission. As apparent from the white isovelocity contours, there is a velocity gradient across the source indicating a presence of a rotating disk. This gradient also shows the potential signature of a warp evident from the change in the position angle (measured counterclockwise from the north direction) from 150\dg\ in the center to 190\dg\ in the outskirts.
We extract multiple position-velocity (PV) slices at position angles between 100\dg\ - 280\dg.
In Fig. \ref{p3:fig:pvslice}, we show a PV slice at a position angle of 150\dg\ which shows a clear gradient in velocity. We see no sign of flat rotation in any of the PV slices.
Furthermore, as \darkgal\ is marginally resolved, the \w50\ is perfectly consistent with the extent of the high S/N ($\sim$4 $\sigma$) emission in all extracted PV slices. 
Hence, we estimate the rotation velocity of \darkgal\ using \w50.

We correct the \w50\ using standard prescriptions for instrumental \citepads{2001A&A...370..765V} and thermal broadening \citepads{1985ApJS...58...67T,2001A&A...370..765V} that have been specifically calibrated for \w50\ measurements on spatially resolved galaxies. We assume the contribution from thermal velocity dispersion of $\sigma^t_{v} = (8 \pm 2)$ \kms, a common value found in dwarf galaxies \citepads{2006AJ....131..363D,2013ApJ...765..136S}. The projected rotational velocity after the above corrections is found to be (10.2 $\pm$ 2.2) \kms.

As we cannot determine \hi\ geometry of \darkgal\ due to insufficient spatial resolution, we employ the optical geometry for the inclination correction of the rotational velocity. However, we note that the optical $PA$ does not seem to be fully aligned\footnote{Beam smearing in \hi\ observations might be shifting the observed \PAkin\ to higher values in the inner parts due to the presence of the warp. Hence, it is possible that \hi\ observations with improved spatial resolution would find \PAkin\ in the central part to be consistent with the optical one.
} with the kinematic position angle (with $\sim$22-62\dg\ difference for \PAkin\ 150-190\dg), making the following estimate potentially uncertain. 
In addition, a warp may introduce a systematic uncertainty in the rotational velocity estimate, which could be biased toward lower values if the outer disk (assuming it dominates the \w50\ measurement) is warped to a lower inclination than the inner disk. However, the presence and magnitude of this effect are difficult to estimate given the limited spatial resolution of the data.
Taking the optically determined inclination of 56\dg\ (Sect. \ref{p3:sec:optical_properties}), we obtain a rotational velocity of \vrot\ = (12.3 $\pm$ 2.8) \kms.

\subsection{Comparison of \hi\ properties between VLA and FAST}

As previously mentioned in Sect. \ref{p3:sec:intro}, the galaxy was initially detected with FAST and reported in X23 as a dark galaxy. Additionally, there is another FAST detection of the source within the FASHI catalog (hereafter FASHI detection; \citeads{2024SCPMA..6719511Z}). In this section, we compare the obtained \hi\ properties of the VLA detection to the two detections from FAST. As X23 reported their \hi\ measurements in the heliocentric rest frame and radio velocity convention, we convert our quantities to the same convention and rest frame for a consistent comparison.

The \hi\ centroid of the X23 (FASHI) detection is 31\arcsec\ (13\arcsec) in the south direction offset from the VLA one. The relatively large offset between the X23 centroid and the VLA one (which is positioned on top of the optical counterpart) is attributed to the centroiding error of the FAST telescope\footnote{As shown for the FASHI catalog in figure 9 of \citeads{2024SCPMA..6719511Z}, the centroiding error of the FAST telescope peaks at $\sim 20$\arcsec\ with a tail towards higher values, so the offset of 31\arcsec\ found in this work is not a significant outlier.} and was likely the reason for the misinterpretation of the galaxy as a dark \hi\ cloud. The systemic velocities determined in X23 (2464.4 \kms) and reported in FASHI (2462.8 \kms) are well consistent with the VLA estimate of 2463.1 \kms.
The total flux of the X23 detection is 424 mJy \kms; 1.6 times higher than the VLA flux of 259 mJy \kms. On the other hand, the flux from the FASHI catalog amounts to only 256 mJy \kms\ (well within 1$\sigma$ of the VLA flux). 
Looking at the global profile, the X23 profile peaks at a higher value of $\sim$12 mJy, while the VLA and the FASHI detections both peak at $\sim$9 mJy.
Furthermore, when comparing the shapes of the X23 and the VLA profiles, the VLA spectrum falls off more quickly toward lower velocities, corresponding to the approaching part of the galaxy. Consequently, the \w50\ is larger for the X23 detection (38.9 \kms) compared to the VLA (28.2 \kms), while the \w50\ of the FASHI detection is between those two with 35.1 \kms. With the same peak value, but a larger \w50, the FASHI detection likely has a more similar shape to the X23 detection, but is scaled toward lower fluxes, in line with the VLA flux.

While there are dissimilarities between all \hi\ measurements of \darkgal, we cannot conclusively determine whether their origin\footnote{Differences in flux measurements are unlikely to result from the VLA’s interferometric short-spacing problem, since the maximum angular extent of the source in the X23 detection ($\sim$6\arcmin) is well below the spatial scales at which the VLA begins to lose sensitivity (10–12\arcmin, set by the 35.5 m minimum baseline in the C configuration).} is due to differences in spatial resolution, column density sensitivity, or overall calibration of the \hi\ data. It is, however, clear that the source is at most marginally resolved in all of these detections at their respective column density sensitivities. Deeper \hi\ data with at least the same resolution as the VLA and column density sensitivities of FAST ($\sim 10^{18}$ cm$^{-2}$) is needed to truly tackle the observed differences between different detections as well as to robustly characterize the source in terms of its kinematic properties and the morphology of the \hi\ disk.

\subsection{Optical properties}
\label{p3:sec:optical_properties}

In this section, we describe our photometric measurements and the derivation of the stellar mass estimate for \darkgal. We note there is a red background source coincident with the galaxy, as seen in the right panel of Fig. \ref{p3:fig:hiopt}, although offset sufficiently from the center to still enable reliable photometry. We mask this source together with foreground stars before estimating the galaxy's geometry using isophotal fitting. We employ the isophotal fitting procedure described in \citetads{2024A&A...692A.217S}. In short, we fit the $i-$band image using the Astropy affiliated package \texttt{photutils} \citepads{larry_bradley_2022_7419741} two times, first time constraining the position of the center, and second time constraining the position angle and the ellipticity. We smoothed the $i-$band image with a 3 pixel smoothing kernel before the fitting to increase the S/N. We take the median value of each parameter within the region outside the FWHM of the smoothed PSF and the surface brightness limit of 27 mag/arcsec$^2$ (below this limit, the geometry is highly influenced by noise in the image). The obtained optical geometry is given in Table \ref{p3:tab:props}, and the fitted model is shown in Fig. \ref{p3:fig:isophot_fit}. 

Due to the faintness and the small size of the source, we prefer to use a fixed aperture photometry instead of a fitted model.
Aperture photometry was performed on full spatial resolution images using the obtained optical geometry and a semi-major axis extent of 12\arcsec. As seen in Fig. \ref{p3:fig:photometry}, the 12\arcsec\ aperture incorporates all the visible flux from the galaxy. 
For confirmation, we have experimented with smaller (6\arcsec\ and 9\arcsec) and larger extents (15\arcsec\ and 18\arcsec), confirming that the obtained magnitudes tend to be fainter in the former case (indicating that some galaxy emission has likely been excluded), and compatible with the chosen extent in the latter (no additional emission was included). For example, a 9\arcsec\ aperture resulted in systematically $\sim0.14$ fainter magnitudes, while a 15\arcsec\ aperture yielded magnitudes consistent within 1.5 $\sigma$.
An elliptical annulus of the same shape with semi-major axes from 18\arcsec\--24\arcsec\ was used for a local sky background subtraction. Errors on the magnitudes are estimated by placing the same aperture in 12 positions without any sources in the image and repeating the same procedure. Standard deviation of these values was used as the error estimate for the obtained magnitude. We measured photometry using the masked images and report the results in Table \ref{p3:tab:props}. We note that the flux within the 12\arcsec\ aperture of an unmasked image (keeping the same background level measured from the masked image) gave fully compatible results within 1$\sigma$ error.

Correcting for Galactic extinction from \citetads{2011ApJ...737..103S} and using the mass-to-light ratio - ($g - r$) color relation from \citetads{2016AJ....152..177H}, we find a stellar mass of $3.2 \times 10^6$ \msun\footnote{During the reviewing process of this paper, \citetads{2025arXiv251224924M} published another detection of the optical counterpart of \darkgal\ idetified using deep optical imaging. The stellar mass estimate obtained in their work is fully consistent with that reported here.}
, reported in Table \ref{p3:tab:props}. Furthermore, if we transfer to the V-band magnitude using the filter transformations from \citetads{2012ApJ...750...99T}, we obtain the V-band luminosity of $L_V = -12.28$. This gives log(\mhi/$L_V$) of 0.94, a value closely comparable to the ones of UDGs ($\sim 0.2 - 1$), and significantly larger than a median of the ALFALFA-SDSS sample ($\sim -0.3$; \citeads{2020AJ....160..271D,2021AJ....162..274L}).


\section{Discussion}\label{p3:sec:discussion}

\subsection{Environment of \darkgal}

We searched for the closest known galaxy to \darkgal\ within a systemic velocity difference of 1000 \kms\ using the NASA/IPAC Extragalactic Database\footnote{\url{https://ned.ipac.caltech.edu/}} (NED). The closest neighbor to \darkgal\ is NGC\,620, with an angular separation of 1.24\dg\ and a nearly identical recessional velocity, of 2507 \kms (separated by only 23 \kms) \citepads{1998A&AS..130..333T}. At a distance of 31.03 Mpc, that corresponds to a projected separation of 671 kpc.
NGC\,620 is included by \citetads{2011AstBu..66....1K} in their isolated nearby galaxy sample ( “Local Orphan Galaxies”, or LOGs).
For a distance of 31.03 Mpc, the stellar mass
of NGC\,620 is $2.7 \times 10^{9}$ \msun \citepads{2019ApJS..244...24L}.
Based on abundance matching, NGC\,620 has a halo mass of 
$1.9 \times 10^{11}$ \msun\ \citepads{2010ApJ...717..379B}.
Such a halo has a virial radius of around 155 kpc (assuming the over-density parameter of $\Delta = 300$), much smaller
than the projected separation of 671 kpc to \darkgal.
Thus, \darkgal\ is very unlikely to be a satellite galaxy or a tidal dwarf, making it highly probable that its properties (e.g. its large gas richness) are intrinsic to the galaxy.

\subsection{Scaling relations}

While initially proposed as a dark galaxy, we have now confirmed the presence of a faint stellar counterpart. Given the low stellar content, we wish to understand if this is a typical dwarf galaxy, simply selected via \hi\ emission and lying near the detection threshold of current optical surveys for its stellar mass and distance, or if it represents an unusual dwarf galaxy, indicative of new populations we might find with advancing \hi\ surveys.

\begin{figure}
    \centering
    \includegraphics[width=0.95\linewidth]{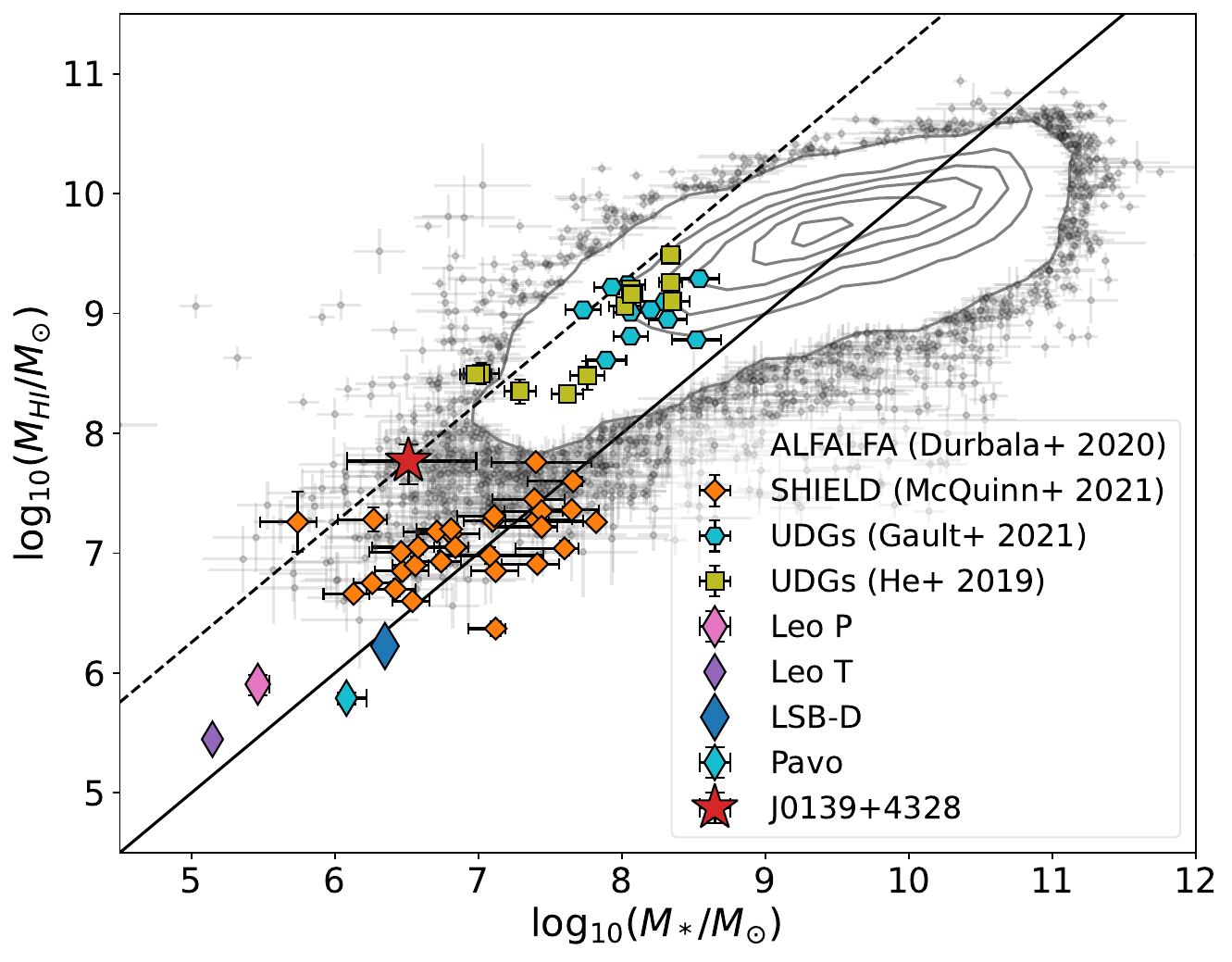}
    \caption{\mhi\ vs \mstar\ relation. Full black line denotes the 1:1 relation (gas-richness of 1), and the dashed line denotes the 1:18 relation (gas-richness of 18).}
    \label{p3:fig:mhi_mstar}
\end{figure}

We first look at the \mhi\ - \mstar\ relation, shown in Fig. \ref{p3:fig:mhi_mstar}. We plot the ALFALFA-SDSS catalog \citepads{2020AJ....160..271D} for context. 
We also plot a subsample of the Survey of \hi\ in Extremely Low-mass Dwarfs (SHIELD; \citeads{2021ApJ...918...23M}), a sample selected from the ALFALFA survey with \hi\ mass \mhi\, $\lesssim 10^{7.2}$ \msun. 
We include two samples of ultra diffuse galaxies as potential analogs of optically faint and gas rich galaxies at higher stellar masses: edge-on \hi -bearing ultra diffuse galaxies (UDGs; \citeads{2019ApJ...880...30H}) and UDGs from \citetads{2021ApJ...909...19G}, both selected from the ALFALFA survey.
We also add Leo P \citepads{2013AJ....146...15G,2024ApJ...976...60M} and Leo T \citepads{2007ApJ...656L..13I,2007ApJ...670..313S,2008MNRAS.384..535R} as local \hi-rich faint dwarf galaxies for context. Finally, we add LSB-D dwarf from the Dorado galaxy group \citepads{2024A&A...690A..69M} and Pavo \citepads{2023ApJ...957L...5J,2025arXiv250606424J}, one of the lowest (baryonic) mass \hi-rich galaxies thus far known outside the Local Group.

\darkgal\ has a stellar mass similar to the SHIELD sample and the LSB-D, but is significantly gas-richer, falling into a regime that has not been well studied before at high resolution. This regime is naturally excluded from SHIELD due to the upper \hi\ mass limit in their selection, and is also excluded from UDG selections due to the small intrinsic sizes of such low stellar mass objects. With gas-richness (\mhi / \mstar) of 18, \darkgal\ is significantly more gas-rich than our local faint dwarfs Leo P and Leo T, and is instead in the regime
of UDG galaxy samples, but is an order of magnitude smaller in mass. It would hence be interesting to explore if \darkgal\ could be a low-mass counterpart to \hi-rich UDGs, that is, having lower surface brightness and larger effective radii when compared to its similar mass dwarfs, using deep optical imaging. Finally, while \darkgal\ is gas-rich for its stellar mass, it is not a significant outlier from the relation traced by the ALFALFA-SDSS catalog, which is naturally biased toward gas-rich systems at these masses.

Next, we look at the Baryonic Tully Fisher Relation (BTFR). The BTFR connects the baryonic mass of the galaxy to the circular velocity of its dark matter halo, and is one of the tightest scaling relations for late-type galaxies. Interestingly, the \hi-rich UDGs have been shown to be systematically offset from this relation, having lower circular velocities compared to galaxies of similar baryonic content \citepads{2019ApJ...883L..33M,2020MNRAS.495.3636M}. We therefore investigate whether \darkgal\ follows this trend and shares dynamical properties with \hi-rich UDGs, or if it aligns more closely with the scaling relations of typical dwarf galaxies.

\begin{figure}
    \centering
    \includegraphics[width=0.95\linewidth]{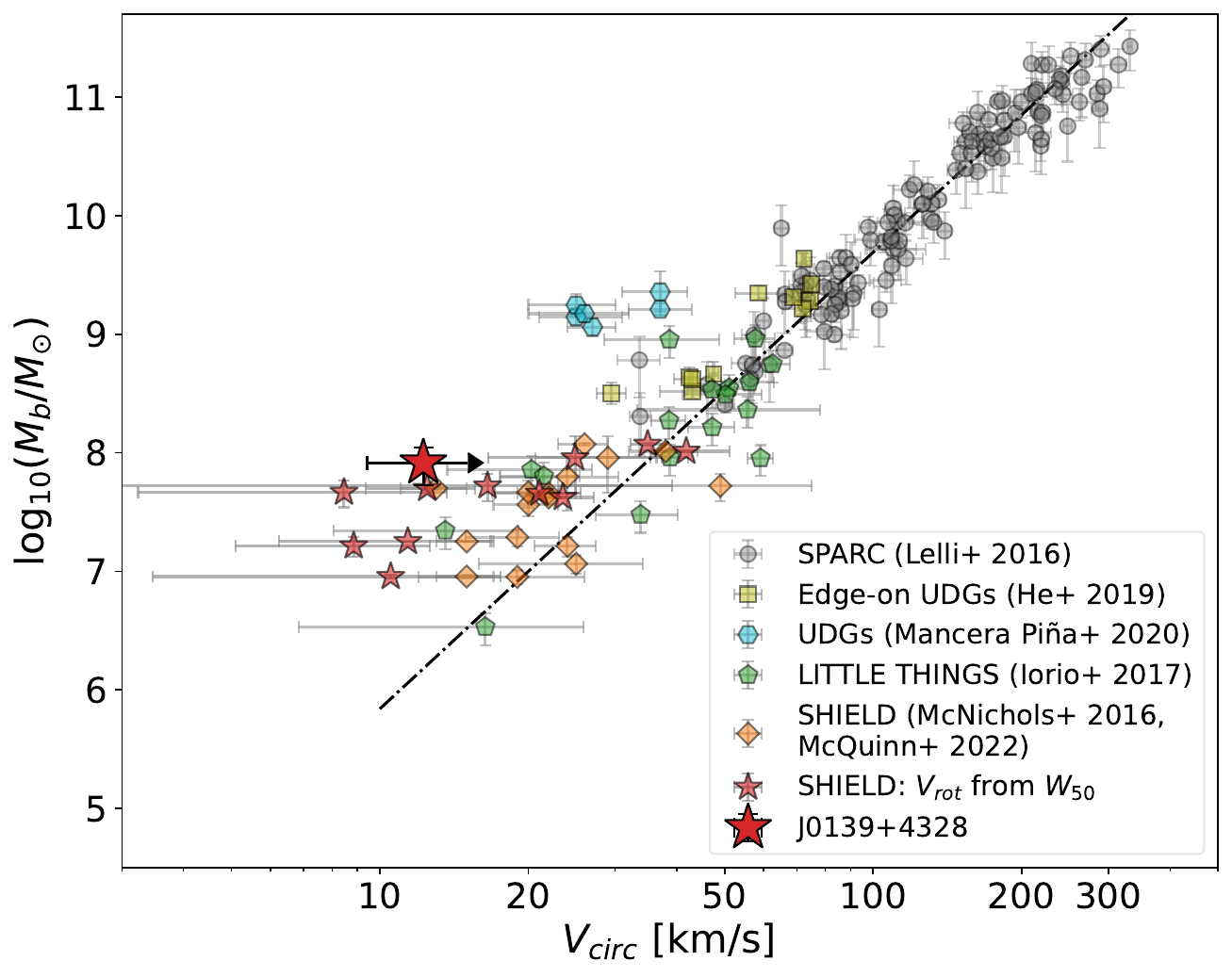}
    \caption{Position of \darkgal\ on the baryonic Tully Fisher relation. Black dash dotted line is the best fit model to the SPARC sample from \citetads{2016AJ....152..157L}.}
    \label{p3:fig:BTFR}
\end{figure}

The BTFR is shown in Fig. \ref{p3:fig:BTFR}. We plot various samples from the literature for context.  
We take 123 galaxies from the SPARC sample \citepads{2016AJ....152..157L} with reliable rotation curves (quality flags $Q=1$ or 2) and inclinations greater than $25$\dg, excluding 3 galaxies that are part of the LITTLE THINGS subsample. The LITTLE THINGS subsample with 17 galaxies comes from \citetads{2017MNRAS.466.4159I} that has a more detailed analysis (based on the 3D modeling of the whole data cube). In addition, we incorporated a subsample of 16 galaxies from the SHIELD survey with published velocity measurements \citepads{2016ApJ...832...89M,2022ApJ...940....8M}. We also include two sets of UDGs: a sample of six systems from \citetads{2019ApJ...883L..33M,2020MNRAS.495.3636M} (including the updated velocity for AGC 114905 from \citeads{2024A&A...689A.344M}), and a sample of 11 edge-on UDGs from \citetads{2019ApJ...880...30H}. 

All the above samples (except the edge-on UDGs from \citeads{2019ApJ...880...30H}) are based on spatially well resolved \hi\ measurements (making them more reliable for tracing the true \vrot). Furthermore, all the resolved \hi\ rotation curves (except the SHIELD subsample from \citetads{2016ApJ...832...89M}) were corrected for the asymmetric drift, thereby obtaining the circular speed of the system \vcirc. 
For a more consistent comparison with the marginally resolved \darkgal\ for which such detailed analysis is impossible, we additionally overplot the same SHIELD galaxies again, this time applying the same procedure for \vrot\ estimation we used for \darkgal\ (see Sect. \ref{p3:sec:HI_properties}; from \w50\ reported in the ALFALFA catalog and with the same inclinations used for their resolved \vrot\ estimates). We note that in 5 galaxies our assumed thermal broadening of 8 \kms\ (together with instrumental broadening) could explain the whole ALFALFA \w50\ measurement, leaving no space for rotation. Hence, we cannot estimate the rotation velocity using our method in these cases and we leave them out of this analysis.

\darkgal\ has a lower rotational velocity compared to spatially resolved measurements of galaxies with similar baryonic mass. However, after applying the same procedure for \vrot\ estimation to the SHIELD sample, we see that many galaxies move toward lower rotational velocities, in the similar regime as \darkgal.
In addition, we are not tracing the flat part of the rotation curve and our inclination correction is potentially uncertain due to the slight misalignment between the kinematic and optical position angles.
Hence, although \darkgal\ exhibits potentially similar properties to UDGs, we cannot discern its exact position in the BTFR based on our data.

\section{Conclusions and future considerations}
\label{p3:sec:conclusions_and_outlook}

In this work, we have presented the follow-up VLA \hi\ observations of \darkgal, a former dark galaxy candidate originally detected with the single-dish FAST telescope (X23). The VLA centroid is 31\arcsec\ offset from the FAST detection and is situated on top of a faint optical counterpart. While \darkgal\ is optically faint, it does not seem to be a significant outlier from the stellar-to-\hi\ mass relation traced by the ALFALFA-SDSS sample nor the baryonic Tully-Fisher relation when the scatter is taken into account (although the kinematic measurement is too uncertain to support strong conclusions). It does, however, populate a parameter space of low stellar mass with high gas richness, that has not been well studied before and could potentially exhibit similar properties to UDGs.

This case highlights the importance of cautious interpretation of the current state-of-the-art single-dish \hi\ data.
In the example of \darkgal, a galaxy of $3\times 10^6$\msun\ in stellar mass detected at a distance of 31 Mpc, it is clear that the FAST \hi\ surveys are outstripping optical surveys in sensitivity to low-mass, low surface brightness, gas-rich galaxies. The claim of \darkgal\ as a dark galaxy highlights the importance of care in interpreting the shallow optical data when comparing to the deep but low-resolution FAST data. This case demonstrates that single-dish \hi\ data require meticulous search for the optical counterpart within the centroiding error of the telescope and even beyond, as also seen for some of the detections from ALFALFA \citepads{2015AJ....149...72C}. Hence, in cases such as the one of \darkgal, interferometric follow-ups might be crucial for robust identification of the optical counterpart.

The identification of an optical counterpart to the FAST \hi\ detection highlights that there are very few truly "dark"
systems for the sensitivies of surveys to date. Most "dark" \hi\ detections are tidal debris and even those often have optical counterparts with deep enough optical data (e.g. the ALFALFA Virgo 7 cloud complex, \citeads{2024ApJ...966L..15J}; or the WALLABY J103508 - 283427 \citeads{2024MNRAS.528.4010O}). As ongoing and upcoming future \hi\ surveys push to fainter sensitivities, a "dark" space may await discovery but requires optical data of a matching sensitivity for robust claims.
The “almost” dark galaxies present an exciting discovery in and of themselves, especially as new \hi\ observations probe to lower column densities, allowing us to explore whether gas disks can persist in more diffuse and tenuous states.

\begin{acknowledgements}
We thank the anonymous referee for valuable comments that helped improve this manuscript. This work has received funding from the European Research Council (ERC) under the Horizon Europe research and innovation programme (Acronym: FLOWS, Grant number: 101096087). KMH acknowledges financial support from the grant CEX2021-001131-S funded by MCIN/AEI/ 10.13039/501100011033 from the coordination of the participation in SKA-SPAIN funded by the Ministry of Science and Innovation (MCIN); and from grant PID2021-123930OB-C21 funded by MCIN/AEI/ 10.13039/501100011033 by “ERDF A way of making Europe” and by the "European Union". KMH acknowledges funding from the ERC under the European Union’s Seventh Framework Programme (FP/2007–2013)/ERC Grant Agreement No. 291531 (‘HIStoryNU’).

JLX and MZ acknowledge the support of the National Key R\&D Program of China No. 2022YFA1602901. This work is also supported by the Youth Innovation Promotion Association of CAS, the National Natural Science Foundation of China (grant No. 11933011), the Central Government Funds for Local Scientific and Technological Development (No. XZ202201YD0020C), and supported by the Open Project Program of the Key Laboratory of FAST, NAOC, Chinese Academy of Sciences.

The National Radio Astronomy Observatory and Green Bank Observatory are facilities of the U.S. National Science Foundation operated under cooperative agreement by Associated Universities, Inc. The Pan-STARRS1 Surveys (PS1) have been made possible through contributions of the Institute for Astronomy, the University of Hawaii, the Pan-STARRS Project Office, the Max-Planck Society and its participating institutes, the Max Planck Institute for Astronomy, Heidelberg and the Max Planck Institute for Extraterrestrial Physics, Garching, The Johns Hopkins University, Durham University, the University of Edinburgh, Queen's University Belfast, the Harvard-Smithsonian Center for Astrophysics, the Las Cumbres Observatory Global Telescope Network Incorporated, the National Central University of Taiwan, the Space Telescope Science Institute, the National Aeronautics and Space Administration under Grant No. NNX08AR22G issued through the Planetary Science Division of the NASA Science Mission Directorate, the National Science Foundation under Grant No. AST-1238877, the University of Maryland, and Eotvos Lorand University (ELTE).

This work made use of Astropy:\footnote{http://www.astropy.org} a community-developed core Python package and an ecosystem of tools and resources for astronomy \citepads{astropy:2013, astropy:2018, astropy:2022}. This research made use of Photutils, an Astropy package for detection and photometry of astronomical sources \citep{larry_bradley_2023_1035865}. This work made use of SpectralCube (DOI: 10.5281/zenodo.591639). This research has made use of the NASA/IPAC Extragalactic Database (NED), which is operated by the Jet Propulsion Laboratory, California Institute of Technology, under contract with the National Aeronautics and Space Administration.

\end{acknowledgements}

\bibliographystyle{aa}
\bibliography{refs}

@ARTICLE{2018A&A...612A..26A,
       author = {{Adams}, Elizabeth A.~K. and {Oosterloo}, Tom A.},
        title = "{Deep neutral hydrogen observations of Leo T with the Westerbork Synthesis Radio Telescope}",
      journal = {\aap},
         year = 2018,
        month = apr,
       volume = {612},
          eid = {A26},
        pages = {A26},
          doi = {10.1051/0004-6361/201732017},
archivePrefix = {arXiv},
       eprint = {1712.06636},
 primaryClass = {astro-ph.GA},
       adsurl = {https://ui.adsabs.harvard.edu/abs/2018A&A...612A..26A}
}

@ARTICLE{2011ApJ...737..103S,
       author = {{Schlafly}, Edward F. and {Finkbeiner}, Douglas P.},
        title = "{Measuring Reddening with Sloan Digital Sky Survey Stellar Spectra and Recalibrating SFD}",
      journal = {\apj},
         year = 2011,
        month = aug,
       volume = {737},
       number = {2},
          eid = {103},
        pages = {103},
          doi = {10.1088/0004-637X/737/2/103},
archivePrefix = {arXiv},
       eprint = {1012.4804},
 primaryClass = {astro-ph.GA},
       adsurl = {https://ui.adsabs.harvard.edu/abs/2011ApJ...737..103S}
}

@ARTICLE{2016AJ....152..177H,
       author = {{Herrmann}, Kimberly A. and {Hunter}, Deidre A. and {Zhang}, Hong-Xin and {Elmegreen}, Bruce G.},
        title = "{Mass-to-light versus Color Relations for Dwarf Irregular Galaxies}",
      journal = {\aj},
         year = 2016,
        month = dec,
       volume = {152},
       number = {6},
          eid = {177},
        pages = {177},
          doi = {10.3847/0004-6256/152/6/177},
archivePrefix = {arXiv},
       eprint = {1701.00144},
 primaryClass = {astro-ph.GA},
       adsurl = {https://ui.adsabs.harvard.edu/abs/2016AJ....152..177H}
}

@ARTICLE{2013A&A...555L...7O,
       author = {{Oosterloo}, T.~A. and {Heald}, G.~H. and {de Blok}, W.~J.~G.},
        title = "{Is GBT 1355+5439 a dark galaxy?}",
      journal = {\aap},
         year = 2013,
        month = jul,
       volume = {555},
          eid = {L7},
        pages = {L7},
          doi = {10.1051/0004-6361/201321965},
archivePrefix = {arXiv},
       eprint = {1306.6148},
 primaryClass = {astro-ph.GA},
       adsurl = {https://ui.adsabs.harvard.edu/abs/2013A&A...555L...7O}
}

@ARTICLE{2023ApJ...944L..40X,
       author = {{Xu}, Jin-Long and {Zhu}, Ming and {Yu}, Naiping and {Zhang}, Chuan-Peng and {Liu}, Xiao-Lan and {Ai}, Mei and {Jiang}, Peng},
        title = "{Discovery of an Isolated Dark Dwarf Galaxy in the Nearby Universe}",
      journal = {\apjl},
         year = 2023,
        month = feb,
       volume = {944},
       number = {2},
          eid = {L40},
        pages = {L40},
          doi = {10.3847/2041-8213/acb932},
archivePrefix = {arXiv},
       eprint = {2302.02646},
 primaryClass = {astro-ph.CO},
       adsurl = {https://ui.adsabs.harvard.edu/abs/2023ApJ...944L..40X}
}

@ARTICLE{2020AJ....159...67K,
       author = {{Kourkchi}, Ehsan and {Courtois}, H{\'e}l{\`e}ne M. and {Graziani}, Romain and {Hoffman}, Yehuda and {Pomar{\`e}de}, Daniel and {Shaya}, Edward J. and {Tully}, R. Brent},
        title = "{Cosmicflows-3: Two Distance-Velocity Calculators}",
      journal = {\aj},
         year = 2020,
        month = feb,
       volume = {159},
       number = {2},
          eid = {67},
        pages = {67},
          doi = {10.3847/1538-3881/ab620e},
archivePrefix = {arXiv},
       eprint = {1912.07214},
 primaryClass = {astro-ph.CO},
       adsurl = {https://ui.adsabs.harvard.edu/abs/2020AJ....159...67K}
}

@ARTICLE{2017ApJ...850..207S,
       author = {{Shaya}, Edward J. and {Tully}, R. Brent and {Hoffman}, Yehuda and {Pomar{\`e}de}, Daniel},
        title = "{Action Dynamics of the Local Supercluster}",
      journal = {\apj},
         year = 2017,
        month = dec,
       volume = {850},
       number = {2},
          eid = {207},
        pages = {207},
          doi = {10.3847/1538-4357/aa9525},
archivePrefix = {arXiv},
       eprint = {1710.08935},
 primaryClass = {astro-ph.CO},
       adsurl = {https://ui.adsabs.harvard.edu/abs/2017ApJ...850..207S}
}

@ARTICLE{2011AstBu..66....1K,
       author = {{Karachentsev}, I.~D. and {Makarov}, D.~I. and {Karachentseva}, V.~E. and {Melnyk}, O.~V.},
        title = "{Catalog of nearby isolated galaxies in the volume z < 0.01}",
      journal = {Astrophysical Bulletin},
         year = 2011,
        month = jan,
       volume = {66},
       number = {1},
        pages = {1-27},
          doi = {10.1134/S1990341311010019},
archivePrefix = {arXiv},
       eprint = {1103.3990},
 primaryClass = {astro-ph.CO},
       adsurl = {https://ui.adsabs.harvard.edu/abs/2011AstBu..66....1K}
}

@ARTICLE{1998A&AS..130..333T,
       author = {{Theureau}, G. and {Bottinelli}, L. and {Coudreau-Durand}, N. and {Gouguenheim}, L. and {Hallet}, N. and {Loulergue}, M. and {Paturel}, G. and {Teerikorpi}, P.},
        title = "{Kinematics of the local universe. VII. New 21-cm line measurements of 2112 galaxies}",
      journal = {\aaps},
         year = 1998,
        month = jun,
       volume = {130},
        pages = {333-339},
          doi = {10.1051/aas:1998416},
       adsurl = {https://ui.adsabs.harvard.edu/abs/1998A&AS..130..333T}
}

@ARTICLE{2010ApJ...717..379B,
       author = {{Behroozi}, Peter S. and {Conroy}, Charlie and {Wechsler}, Risa H.},
        title = "{A Comprehensive Analysis of Uncertainties Affecting the Stellar Mass-Halo Mass Relation for 0 < z < 4}",
      journal = {\apj},
         year = 2010,
        month = jul,
       volume = {717},
       number = {1},
        pages = {379-403},
          doi = {10.1088/0004-637X/717/1/379},
archivePrefix = {arXiv},
       eprint = {1001.0015},
 primaryClass = {astro-ph.CO},
       adsurl = {https://ui.adsabs.harvard.edu/abs/2010ApJ...717..379B}
}

@ARTICLE{2019ApJS..244...24L,
       author = {{Leroy}, Adam K. and {Sandstrom}, Karin M. and {Lang}, Dustin and {Lewis}, Alexia and {Salim}, Samir and {Behrens}, Erica A. and {Chastenet}, J{\'e}r{\'e}my and {Chiang}, I-Da and {Gallagher}, Molly J. and {Kessler}, Sarah and {Utomo}, Dyas},
        title = "{A z = 0 Multiwavelength Galaxy Synthesis. I. A WISE and GALEX Atlas of Local Galaxies}",
      journal = {\apjs},
         year = 2019,
        month = oct,
       volume = {244},
       number = {2},
          eid = {24},
        pages = {24},
          doi = {10.3847/1538-4365/ab3925},
archivePrefix = {arXiv},
       eprint = {1910.13470},
 primaryClass = {astro-ph.GA},
       adsurl = {https://ui.adsabs.harvard.edu/abs/2019ApJS..244...24L}
}

@ARTICLE{2016arXiv161205560C,
       author = {{Chambers}, K.~C. and {Magnier}, E.~A. and {Metcalfe}, N. and {Flewelling}, H.~A. and {Huber}, M.~E. and {Waters}, C.~Z. and {Denneau}, L. and {Draper}, P.~W. and {Farrow}, D. and {Finkbeiner}, D.~P. and {Holmberg}, C. and {Koppenhoefer}, J. and {Price}, P.~A. and {Rest}, A. and {Saglia}, R.~P. and {Schlafly}, E.~F. and {Smartt}, S.~J. and {Sweeney}, W. and {Wainscoat}, R.~J. and {Burgett}, W.~S. and {Chastel}, S. and {Grav}, T. and {Heasley}, J.~N. and {Hodapp}, K.~W. and {Jedicke}, R. and {Kaiser}, N. and {Kudritzki}, R. -P. and {Luppino}, G.~A. and {Lupton}, R.~H. and {Monet}, D.~G. and {Morgan}, J.~S. and {Onaka}, P.~M. and {Shiao}, B. and {Stubbs}, C.~W. and {Tonry}, J.~L. and {White}, R. and {Ba{\~n}ados}, E. and {Bell}, E.~F. and {Bender}, R. and {Bernard}, E.~J. and {Boegner}, M. and {Boffi}, F. and {Botticella}, M.~T. and {Calamida}, A. and {Casertano}, S. and {Chen}, W. -P. and {Chen}, X. and {Cole}, S. and {Deacon}, N. and {Frenk}, C. and {Fitzsimmons}, A. and {Gezari}, S. and {Gibbs}, V. and {Goessl}, C. and {Goggia}, T. and {Gourgue}, R. and {Goldman}, B. and {Grant}, P. and {Grebel}, E.~K. and {Hambly}, N.~C. and {Hasinger}, G. and {Heavens}, A.~F. and {Heckman}, T.~M. and {Henderson}, R. and {Henning}, T. and {Holman}, M. and {Hopp}, U. and {Ip}, W. -H. and {Isani}, S. and {Jackson}, M. and {Keyes}, C.~D. and {Koekemoer}, A.~M. and {Kotak}, R. and {Le}, D. and {Liska}, D. and {Long}, K.~S. and {Lucey}, J.~R. and {Liu}, M. and {Martin}, N.~F. and {Masci}, G. and {McLean}, B. and {Mindel}, E. and {Misra}, P. and {Morganson}, E. and {Murphy}, D.~N.~A. and {Obaika}, A. and {Narayan}, G. and {Nieto-Santisteban}, M.~A. and {Norberg}, P. and {Peacock}, J.~A. and {Pier}, E.~A. and {Postman}, M. and {Primak}, N. and {Rae}, C. and {Rai}, A. and {Riess}, A. and {Riffeser}, A. and {Rix}, H.~W. and {R{\"o}ser}, S. and {Russel}, R. and {Rutz}, L. and {Schilbach}, E. and {Schultz}, A.~S.~B. and {Scolnic}, D. and {Strolger}, L. and {Szalay}, A. and {Seitz}, S. and {Small}, E. and {Smith}, K.~W. and {Soderblom}, D.~R. and {Taylor}, P. and {Thomson}, R. and {Taylor}, A.~N. and {Thakar}, A.~R. and {Thiel}, J. and {Thilker}, D. and {Unger}, D. and {Urata}, Y. and {Valenti}, J. and {Wagner}, J. and {Walder}, T. and {Walter}, F. and {Watters}, S.~P. and {Werner}, S. and {Wood-Vasey}, W.~M. and {Wyse}, R.},
        title = "{The Pan-STARRS1 Surveys}",
      journal = {arXiv e-prints},
         year = 2016,
        month = dec,
          eid = {arXiv:1612.05560},
        pages = {arXiv:1612.05560},
          doi = {10.48550/arXiv.1612.05560},
archivePrefix = {arXiv},
       eprint = {1612.05560},
 primaryClass = {astro-ph.IM},
       adsurl = {https://ui.adsabs.harvard.edu/abs/2016arXiv161205560C}
}

@ARTICLE{2024A&A...692A.217S,
       author = {{{\v{S}}iljeg}, B. and {Adams}, E.~A.~K. and {Fraternali}, F. and {Hess}, K.~M. and {Oosterloo}, T.~A. and {Marasco}, A. and {Adebahr}, B. and {D{\'e}nes}, H. and {Garrido}, J. and {Lucero}, D.~M. and {Mancera Pi{\~n}a}, P.~E. and {Moss}, V.~A. and {Parra-Roy{\'o}n}, M. and {Ponomareva}, A.~A. and {S{\'a}nchez-Exp{\'o}sito}, S. and {van der Hulst}, J.~M.},
        title = "{Photometry and kinematics of dwarf galaxies from the Apertif H I survey}",
      journal = {\aap},
         year = 2024,
        month = dec,
       volume = {692},
          eid = {A217},
        pages = {A217},
          doi = {10.1051/0004-6361/202449923},
archivePrefix = {arXiv},
       eprint = {2409.18825},
 primaryClass = {astro-ph.GA},
       adsurl = {https://ui.adsabs.harvard.edu/abs/2024A&A...692A.217S}
}

@software{larry_bradley_2022_7419741,
  author       = {Larry Bradley and
                  Brigitta Sipőcz and
                  Thomas Robitaille and
                  Erik Tollerud and
                  Zé Vinícius and
                  Christoph Deil and
                  Kyle Barbary and
                  Tom J Wilson and
                  Ivo Busko and
                  Axel Donath and
                  Hans Moritz Günther and
                  Mihai Cara and
                  P. L. Lim and
                  Sebastian Meßlinger and
                  Simon Conseil and
                  Azalee Bostroem and
                  Michael Droettboom and
                  E. M. Bray and
                  Lars Andersen Bratholm and
                  Geert Barentsen and
                  Matt Craig and
                  Adam Ginsburg and
                  Shivangee Rathi and
                  Sergio Pascual and
                  Gabriel Perren and
                  Iskren Y. Georgiev and
                  Miguel de Val-Borro and
                  Wolfgang Kerzendorf and
                  Yoonsoo P. Bach and
                  Bruno Quint},
  title        = {astropy/photutils: 1.6.0},
  month        = dec,
  year         = 2022,
  publisher    = {Zenodo},
  version      = {1.6.0},
  doi          = {10.5281/zenodo.7419741},
  url          = {https://doi.org/10.5281/zenodo.7419741}
}

@ARTICLE{2016AJ....152..157L,
       author = {{Lelli}, Federico and {McGaugh}, Stacy S. and {Schombert}, James M.},
        title = "{SPARC: Mass Models for 175 Disk Galaxies with Spitzer Photometry and Accurate Rotation Curves}",
      journal = {\aj},
         year = 2016,
        month = dec,
       volume = {152},
       number = {6},
          eid = {157},
        pages = {157},
          doi = {10.3847/0004-6256/152/6/157},
archivePrefix = {arXiv},
       eprint = {1606.09251},
 primaryClass = {astro-ph.GA},
       adsurl = {https://ui.adsabs.harvard.edu/abs/2016AJ....152..157L}
}

@ARTICLE{2001A&A...370..765V,
       author = {{Verheijen}, M.~A.~W. and {Sancisi}, R.},
        title = "{The Ursa Major cluster of galaxies. IV. HI synthesis observations}",
      journal = {\aap},
         year = 2001,
        month = may,
       volume = {370},
        pages = {765-867},
          doi = {10.1051/0004-6361:20010090},
archivePrefix = {arXiv},
       eprint = {astro-ph/0101404},
 primaryClass = {astro-ph},
       adsurl = {https://ui.adsabs.harvard.edu/abs/2001A&A...370..765V}
}

@ARTICLE{1985ApJS...58...67T,
       author = {{Tully}, R.~B. and {Fouque}, P.},
        title = "{The extragalactic distance scale. I - Corrections to fundamental observables.}",
      journal = {\apjs},
         year = 1985,
        month = may,
       volume = {58},
        pages = {67-80},
          doi = {10.1086/191029},
       adsurl = {https://ui.adsabs.harvard.edu/abs/1985ApJS...58...67T}
}

@ARTICLE{2005AJ....130.2598G,
       author = {{Giovanelli}, Riccardo and {Haynes}, Martha P. and {Kent}, Brian R. and {Perillat}, Philip and {Saintonge}, Amelie and {Brosch}, Noah and {Catinella}, Barbara and {Hoffman}, G. Lyle and {Stierwalt}, Sabrina and {Spekkens}, Kristine and {Lerner}, Mikael S. and {Masters}, Karen L. and {Momjian}, Emmanuel and {Rosenberg}, Jessica L. and {Springob}, Christopher M. and {Boselli}, Alessandro and {Charmandaris}, Vassilis and {Darling}, Jeremy K. and {Davies}, Jonathan and {Garcia Lambas}, Diego and {Gavazzi}, Giuseppe and {Giovanardi}, Carlo and {Hardy}, Eduardo and {Hunt}, Leslie K. and {Iovino}, Angela and {Karachentsev}, Igor D. and {Karachentseva}, Valentina E. and {Koopmann}, Rebecca A. and {Marinoni}, Christian and {Minchin}, Robert and {Muller}, Erik and {Putman}, Mary and {Pantoja}, Carmen and {Salzer}, John J. and {Scodeggio}, Marco and {Skillman}, Evan and {Solanes}, Jose M. and {Valotto}, Carlos and {van Driel}, Wim and {van Zee}, Liese},
        title = "{The Arecibo Legacy Fast ALFA Survey. I. Science Goals, Survey Design, and Strategy}",
      journal = {\aj},
         year = 2005,
        month = dec,
       volume = {130},
       number = {6},
        pages = {2598-2612},
          doi = {10.1086/497431},
archivePrefix = {arXiv},
       eprint = {astro-ph/0508301},
 primaryClass = {astro-ph},
       adsurl = {https://ui.adsabs.harvard.edu/abs/2005AJ....130.2598G}
}

@ARTICLE{2021ApJ...918...23M,
       author = {{McQuinn}, Kristen B.~W. and {Telidevara}, Anjana K. and {Fuson}, Jackson and {Adams}, Elizabeth A.~K. and {Cannon}, John M. and {Skillman}, Evan D. and {Dolphin}, Andrew E. and {Haynes}, Martha P. and {Rhode}, Katherine L. and {Salzer}, John. J. and {Giovanelli}, Riccardo and {Gordon}, Alex J.~R.},
        title = "{Galaxy Properties at the Faint End of the H I Mass Function}",
      journal = {\apj},
         year = 2021,
        month = sep,
       volume = {918},
       number = {1},
          eid = {23},
        pages = {23},
          doi = {10.3847/1538-4357/ac03ae},
archivePrefix = {arXiv},
       eprint = {2105.05100},
 primaryClass = {astro-ph.GA},
       adsurl = {https://ui.adsabs.harvard.edu/abs/2021ApJ...918...23M}
}

@ARTICLE{2019ApJ...880...30H,
       author = {{He}, Min and {Wu}, Hong and {Du}, Wei and {Wicker}, James and {Zhao}, Pingsong and {Lei}, Fengjie and {Liu}, Jifeng},
        title = "{Edge-on H I-bearing Ultra-diffuse Galaxy Candidates in the 40\% ALFALFA Catalog}",
      journal = {\apj},
         year = 2019,
        month = jul,
       volume = {880},
       number = {1},
          eid = {30},
        pages = {30},
          doi = {10.3847/1538-4357/ab2710},
archivePrefix = {arXiv},
       eprint = {1907.10438},
 primaryClass = {astro-ph.GA},
       adsurl = {https://ui.adsabs.harvard.edu/abs/2019ApJ...880...30H}
}

@ARTICLE{2020AJ....160..271D,
       author = {{Durbala}, Adriana and {Finn}, Rose A. and {Crone Odekon}, Mary and {Haynes}, Martha P. and {Koopmann}, Rebecca A. and {O'Donoghue}, Aileen A.},
        title = "{The ALFALFA-SDSS Galaxy Catalog}",
      journal = {\aj},
         year = 2020,
        month = dec,
       volume = {160},
       number = {6},
          eid = {271},
        pages = {271},
          doi = {10.3847/1538-3881/abc018},
archivePrefix = {arXiv},
       eprint = {2011.02588},
 primaryClass = {astro-ph.GA},
       adsurl = {https://ui.adsabs.harvard.edu/abs/2020AJ....160..271D}
}

@ARTICLE{2021ApJ...909...19G,
       author = {{Gault}, Lexi and {Leisman}, Lukas and {Adams}, Elizabeth A.~K. and {Mancera Pi{\~n}a}, Pavel E. and {Reiter}, Kameron and {Smith}, Nicholas and {Battipaglia}, Michael and {Cannon}, John M. and {Fraternali}, Filippo and {Haynes}, Martha P. and {McAllan}, Elizabeth and {Pagel}, Hannah J. and {Rhode}, Katherine L. and {Salzer}, John J. and {Singer}, Quinton},
        title = "{VLA Imaging of H I-bearing Ultra-diffuse Galaxies from the ALFALFA Survey}",
      journal = {\apj},
         year = 2021,
        month = mar,
       volume = {909},
       number = {1},
          eid = {19},
        pages = {19},
          doi = {10.3847/1538-4357/abd79d},
archivePrefix = {arXiv},
       eprint = {2101.01753},
 primaryClass = {astro-ph.GA},
       adsurl = {https://ui.adsabs.harvard.edu/abs/2021ApJ...909...19G}
}

@ARTICLE{2019ApJ...883L..33M,
       author = {{Mancera Pi{\~n}a}, Pavel E. and {Fraternali}, Filippo and {Adams}, Elizabeth A.~K. and {Marasco}, Antonino and {Oosterloo}, Tom and {Oman}, Kyle A. and {Leisman}, Lukas and {di Teodoro}, Enrico M. and {Posti}, Lorenzo and {Battipaglia}, Michael and {Cannon}, John M. and {Gault}, Lexi and {Haynes}, Martha P. and {Janowiecki}, Steven and {McAllan}, Elizabeth and {Pagel}, Hannah J. and {Reiter}, Kameron and {Rhode}, Katherine L. and {Salzer}, John J. and {Smith}, Nicholas J.},
        title = "{Off the Baryonic Tully-Fisher Relation: A Population of Baryon-dominated Ultra-diffuse Galaxies}",
      journal = {\apjl},
         year = 2019,
        month = oct,
       volume = {883},
       number = {2},
          eid = {L33},
        pages = {L33},
          doi = {10.3847/2041-8213/ab40c7},
archivePrefix = {arXiv},
       eprint = {1909.01363},
 primaryClass = {astro-ph.GA},
       adsurl = {https://ui.adsabs.harvard.edu/abs/2019ApJ...883L..33M}
}

@ARTICLE{2020MNRAS.495.3636M,
       author = {{Mancera Pi{\~n}a}, Pavel E. and {Fraternali}, Filippo and {Oman}, Kyle A. and {Adams}, Elizabeth A.~K. and {Bacchini}, Cecilia and {Marasco}, Antonino and {Oosterloo}, Tom and {Pezzulli}, Gabriele and {Posti}, Lorenzo and {Leisman}, Lukas and {Cannon}, John M. and {di Teodoro}, Enrico M. and {Gault}, Lexi and {Haynes}, Martha P. and {Reiter}, Kameron and {Rhode}, Katherine L. and {Salzer}, John J. and {Smith}, Nicholas J.},
        title = "{Robust H I kinematics of gas-rich ultra-diffuse galaxies: hints of a weak-feedback formation scenario}",
      journal = {\mnras},
         year = 2020,
        month = jul,
       volume = {495},
       number = {4},
        pages = {3636-3655},
          doi = {10.1093/mnras/staa1256},
archivePrefix = {arXiv},
       eprint = {2004.14392},
 primaryClass = {astro-ph.GA},
       adsurl = {https://ui.adsabs.harvard.edu/abs/2020MNRAS.495.3636M}
}

@ARTICLE{2017MNRAS.466.4159I,
       author = {{Iorio}, G. and {Fraternali}, F. and {Nipoti}, C. and {Di Teodoro}, E. and {Read}, J.~I. and {Battaglia}, G.},
        title = "{LITTLE THINGS in 3D: robust determination of the circular velocity of dwarf irregular galaxies}",
      journal = {\mnras},
         year = 2017,
        month = apr,
       volume = {466},
       number = {4},
        pages = {4159-4192},
          doi = {10.1093/mnras/stw3285},
archivePrefix = {arXiv},
       eprint = {1611.03865},
 primaryClass = {astro-ph.GA},
       adsurl = {https://ui.adsabs.harvard.edu/abs/2017MNRAS.466.4159I}
}

@ARTICLE{2016ApJ...832...89M,
       author = {{McNichols}, Andrew T. and {Teich}, Yaron G. and {Nims}, Elise and {Cannon}, John M. and {Adams}, Elizabeth A.~K. and {Bernstein-Cooper}, Elijah Z. and {Giovanelli}, Riccardo and {Haynes}, Martha P. and {J{\'o}zsa}, Gyula I.~G. and {McQuinn}, Kristen B.~W. and {Salzer}, John J. and {Skillman}, Evan D. and {Warren}, Steven R. and {Dolphin}, Andrew and {Elson}, E.~C. and {Haurberg}, Nathalie and {Ott}, J{\"u}rgen and {Saintonge}, Amelie and {Cave}, Ian and {Hagen}, Cedric and {Huang}, Shan and {Janowiecki}, Steven and {Marshall}, Melissa V. and {Thomann}, Clara M. and {Van Sistine}, Angela},
        title = "{SHIELD: Neutral Gas Kinematics and Dynamics}",
      journal = {\apj},
         year = 2016,
        month = nov,
       volume = {832},
       number = {1},
          eid = {89},
        pages = {89},
          doi = {10.3847/0004-637X/832/1/89},
archivePrefix = {arXiv},
       eprint = {1609.05376},
 primaryClass = {astro-ph.GA},
       adsurl = {https://ui.adsabs.harvard.edu/abs/2016ApJ...832...89M}
}

@ARTICLE{2024SCPMA..6719511Z,
       author = {{Zhang}, Chuan-Peng and {Zhu}, Ming and {Jiang}, Peng and {Cheng}, Cheng and {Wang}, Jing and {Wang}, Jie and {Xu}, Jin-Long and {Liu}, Xiao-Lan and {Yu}, Nai-Ping and {Qian}, Lei and {Yu}, Haiyang and {Ai}, Mei and {Jing}, Yingjie and {Xu}, Chen and {Liu}, Ziming and {Guan}, Xin and {Sun}, Chun and {Yang}, Qingliang and {Huang}, Menglin and {Hao}, Qiaoli and {FAST Collaboration}},
        title = "{The FAST all sky H I survey (FASHI): The first release of catalog}",
      journal = {Science China Physics, Mechanics, and Astronomy},
         year = 2024,
        month = jan,
       volume = {67},
       number = {1},
          eid = {219511},
        pages = {219511},
          doi = {10.1007/s11433-023-2219-7},
archivePrefix = {arXiv},
       eprint = {2312.06097},
 primaryClass = {astro-ph.GA},
       adsurl = {https://ui.adsabs.harvard.edu/abs/2024SCPMA..6719511Z}
}

@ARTICLE{2024ApJ...962..129L,
       author = {{Lee}, Gain and {Hwang}, Ho Seong and {Lee}, Jaehyun and {Shin}, Jihye and {Song}, Hyunmi},
        title = "{Understanding the Formation and Evolution of Dark Galaxies in a Simulated Universe}",
      journal = {\apj},
         year = 2024,
        month = feb,
       volume = {962},
       number = {2},
          eid = {129},
        pages = {129},
          doi = {10.3847/1538-4357/ad1e5d},
archivePrefix = {arXiv},
       eprint = {2401.07007},
 primaryClass = {astro-ph.GA},
       adsurl = {https://ui.adsabs.harvard.edu/abs/2024ApJ...962..129L}
}

@ARTICLE{2021AJ....162..274L,
       author = {{Leisman}, Lukas and {Rhode}, Katherine L. and {Ball}, Catherine and {Pagel}, Hannah J. and {Cannon}, John M. and {Salzer}, John J. and {Janowiecki}, Steven and {Janesh}, William F. and {J{\'o}zsa}, Gyula I.~G. and {Giovanelli}, Riccardo and {Haynes}, Martha P. and {Adams}, Elizabeth A.~K. and {Gray}, Laurin and {Smith}, Nicholas J.},
        title = "{The ALFALFA Almost Dark Galaxy AGC 229101: A 2 Billion Solar Mass H I Cloud with a Very Low Surface Brightness Optical Counterpart}",
      journal = {\aj},
         year = 2021,
        month = dec,
       volume = {162},
       number = {6},
          eid = {274},
        pages = {274},
          doi = {10.3847/1538-3881/ac2a38},
archivePrefix = {arXiv},
       eprint = {2109.12139},
 primaryClass = {astro-ph.GA},
       adsurl = {https://ui.adsabs.harvard.edu/abs/2021AJ....162..274L}
}

@ARTICLE{2019MNRAS.490..566J,
       author = {{Janowiecki}, Steven and {Jones}, Michael G. and {Leisman}, Lukas and {Webb}, Andrew},
        title = "{The environment of H I-bearing ultra-diffuse galaxies in the ALFALFA survey}",
      journal = {\mnras},
         year = 2019,
        month = nov,
       volume = {490},
       number = {1},
        pages = {566-577},
          doi = {10.1093/mnras/stz1868},
archivePrefix = {arXiv},
       eprint = {1906.11543},
 primaryClass = {astro-ph.GA},
       adsurl = {https://ui.adsabs.harvard.edu/abs/2019MNRAS.490..566J}
}

@ARTICLE{2017ApJ...842..133L,
       author = {{Leisman}, Lukas and {Haynes}, Martha P. and {Janowiecki}, Steven and {Hallenbeck}, Gregory and {J{\'o}zsa}, Gyula and {Giovanelli}, Riccardo and {Adams}, Elizabeth A.~K. and {Bernal Neira}, David and {Cannon}, John M. and {Janesh}, William F. and {Rhode}, Katherine L. and {Salzer}, John J.},
        title = "{(Almost) Dark Galaxies in the ALFALFA Survey: Isolated H I-bearing Ultra-diffuse Galaxies}",
      journal = {\apj},
         year = 2017,
        month = jun,
       volume = {842},
       number = {2},
          eid = {133},
        pages = {133},
          doi = {10.3847/1538-4357/aa7575},
archivePrefix = {arXiv},
       eprint = {1703.05293},
 primaryClass = {astro-ph.GA},
       adsurl = {https://ui.adsabs.harvard.edu/abs/2017ApJ...842..133L}
}

@ARTICLE{1989ApJ...346L...5G,
       author = {{Giovanelli}, Riccardo and {Haynes}, Martha P.},
        title = "{A Protogalaxy in the Local Supercluster}",
      journal = {\apjl},
         year = 1989,
        month = nov,
       volume = {346},
        pages = {L5},
          doi = {10.1086/185565},
       adsurl = {https://ui.adsabs.harvard.edu/abs/1989ApJ...346L...5G}
}

@ARTICLE{2025SciA...11S4057L,
       author = {{Liu}, Xiao-Lan and {Xu}, Jin-Long and {Jiang}, Peng and {Zhu}, Ming and {Zhang}, Chuan-Peng and {Yu}, Naiping and {Xu}, Ye and {Guan}, Xin and {Wang}, Jun-Jie},
        title = "{Discovery of a high-velocity cloud of the Milky Way as a potential dark galaxy}",
      journal = {Science Advances},
         year = 2025,
        month = apr,
       volume = {11},
       number = {16},
          eid = {eads4057},
        pages = {eads4057},
          doi = {10.1126/sciadv.ads4057},
archivePrefix = {arXiv},
       eprint = {2504.09419},
 primaryClass = {astro-ph.GA},
       adsurl = {https://ui.adsabs.harvard.edu/abs/2025SciA...11S4057L}
}

@ARTICLE{2020A&A...642L..10B,
       author = {{B{\'\i}lek}, Michal and {M{\"u}ller}, Oliver and {Vudragovi{\'c}}, Ana and {Taylor}, Rhys},
        title = "{Deep optical imaging of the dark galaxy candidate AGESVC1 282}",
      journal = {\aap},
         year = 2020,
        month = oct,
       volume = {642},
          eid = {L10},
        pages = {L10},
          doi = {10.1051/0004-6361/202039174},
archivePrefix = {arXiv},
       eprint = {2009.11300},
 primaryClass = {astro-ph.GA},
       adsurl = {https://ui.adsabs.harvard.edu/abs/2020A&A...642L..10B}
}

@ARTICLE{2012MNRAS.423..787T,
       author = {{Taylor}, R. and {Davies}, J.~I. and {Auld}, R. and {Minchin}, R.~F.},
        title = "{The Arecibo Galaxy Environment Survey - V. The Virgo cluster (I)}",
      journal = {\mnras},
         year = 2012,
        month = jun,
       volume = {423},
       number = {1},
        pages = {787-810},
          doi = {10.1111/j.1365-2966.2012.20914.x},
archivePrefix = {arXiv},
       eprint = {1203.3094},
 primaryClass = {astro-ph.GA},
       adsurl = {https://ui.adsabs.harvard.edu/abs/2012MNRAS.423..787T}
}

@ARTICLE{2013MNRAS.428..459T,
       author = {{Taylor}, R. and {Davies}, J.~I. and {Auld}, R. and {Minchin}, R.~F. and {Smith}, R.},
        title = "{The Arecibo Galaxy Environment Survey - VI. The Virgo cluster (II)}",
      journal = {\mnras},
         year = 2013,
        month = jan,
       volume = {428},
       number = {1},
        pages = {459-469},
          doi = {10.1093/mnras/sts042},
archivePrefix = {arXiv},
       eprint = {1209.4338},
 primaryClass = {astro-ph.GA},
       adsurl = {https://ui.adsabs.harvard.edu/abs/2013MNRAS.428..459T}
}

@ARTICLE{2012AJ....144..159M,
       author = {{Matsuoka}, Y. and {Ienaka}, N. and {Oyabu}, S. and {Wada}, K. and {Takino}, S.},
        title = "{Updated Analysis of a ``Dark'' Galaxy and Its Blue Companion in the Virgo Cloud H I 1225 + 01}",
      journal = {\aj},
         year = 2012,
        month = dec,
       volume = {144},
       number = {6},
          eid = {159},
        pages = {159},
          doi = {10.1088/0004-6256/144/6/159},
archivePrefix = {arXiv},
       eprint = {1210.0073},
 primaryClass = {astro-ph.CO},
       adsurl = {https://ui.adsabs.harvard.edu/abs/2012AJ....144..159M}
}

@ARTICLE{1995AJ....109.2415C,
       author = {{Chengalur}, Jayaram N. and {Giovanelli}, Riccardo and {Haynes}, Martha P.},
        title = "{VLA Observations of HI 1225+01}",
      journal = {\aj},
         year = 1995,
        month = jun,
       volume = {109},
        pages = {2415},
          doi = {10.1086/117460},
       adsurl = {https://ui.adsabs.harvard.edu/abs/1995AJ....109.2415C}
}

@ARTICLE{2025arXiv250504299O,
       author = {{O'Beirne}, T. and {Staveley-Smith}, L. and {Kilborn}, V.~A. and {Wong}, O.~I. and {Westmeier}, T. and {Cluver}, M.~E. and {Bekki}, K. and {Deg}, N. and {D{\'e}nes}, H. and {For}, B. -Q. and {Lee-Waddell}, K. and {Murugeshan}, C. and {Oman}, K. and {Rhee}, J. and {Shen}, A.~X. and {Taylor}, E.~N.},
        title = "{WALLABY pilot survey: properties of HI-selected dark sources and low surface brightness galaxies}",
      journal = {arXiv e-prints},
         year = 2025,
        month = may,
          eid = {arXiv:2505.04299},
        pages = {arXiv:2505.04299},
          doi = {10.48550/arXiv.2505.04299},
archivePrefix = {arXiv},
       eprint = {2505.04299},
 primaryClass = {astro-ph.GA},
       adsurl = {https://ui.adsabs.harvard.edu/abs/2025arXiv250504299O}
}

@ARTICLE{2022ApJ...940....8M,
       author = {{McQuinn}, Kristen. B.~W. and {Adams}, Elizabeth A.~K. and {Cannon}, John M. and {Fuson}, Jackson and {Skillman}, Evan D. and {Brooks}, Alyson and {Rhode}, Katherine L. and {Haynes}, Martha P. and {Inoue}, John L. and {Marine}, Joshua and {Salzer}, John. J. and {Talluri}, Anjana K.},
        title = "{The Turndown of the Baryonic Tully-Fisher Relation and Changing Baryon Fraction at Low Galaxy Masses}",
      journal = {\apj},
         year = 2022,
        month = nov,
       volume = {940},
       number = {1},
          eid = {8},
        pages = {8},
          doi = {10.3847/1538-4357/ac9285},
archivePrefix = {arXiv},
       eprint = {2203.10105},
 primaryClass = {astro-ph.GA},
       adsurl = {https://ui.adsabs.harvard.edu/abs/2022ApJ...940....8M}
}

@ARTICLE{2015AJ....149...72C,
       author = {{Cannon}, John M. and {Martinkus}, Charlotte P. and {Leisman}, Lukas and {Haynes}, Martha P. and {Adams}, Elizabeth A.~K. and {Giovanelli}, Riccardo and {Hallenbeck}, Gregory and {Janowiecki}, Steven and {Jones}, Michael and {J{\'o}zsa}, Gyula I.~G. and {Koopmann}, Rebecca A. and {Nichols}, Nathan and {Papastergis}, Emmanouil and {Rhode}, Katherine L. and {Salzer}, John J. and {Troischt}, Parker},
        title = "{The Alfalfa {\textquotedblleft}Almost Darks{\textquotedblright} Campaign: Pilot VLA HI Observations of Five High Mass-To-Light Ratio Systems}",
      journal = {\aj},
         year = 2015,
        month = feb,
       volume = {149},
       number = {2},
          eid = {72},
        pages = {72},
          doi = {10.1088/0004-6256/149/2/72},
archivePrefix = {arXiv},
       eprint = {1412.3018},
 primaryClass = {astro-ph.GA},
       adsurl = {https://ui.adsabs.harvard.edu/abs/2015AJ....149...72C}
}

@ARTICLE{2024ApJ...966L..15J,
       author = {{Jones}, Michael G. and {Janowiecki}, Steven and {Dey}, Swapnaneel and {Sand}, David J. and {Bennet}, Paul and {Crnojevi{\'c}}, Denija and {Fielder}, Catherine E. and {Karunakaran}, Ananthan and {Kent}, Brian R. and {Mazziotti}, Nicolas and {Mutlu-Pakdil}, Bur{\c{c}}in and {Spekkens}, Kristine},
        title = "{Dark No More: The Low-luminosity Stellar Counterpart of a Dark Cloud in the Virgo Cluster}",
      journal = {\apjl},
         year = 2024,
        month = may,
       volume = {966},
       number = {1},
          eid = {L15},
        pages = {L15},
          doi = {10.3847/2041-8213/ad3ef5},
archivePrefix = {arXiv},
       eprint = {2402.14909},
 primaryClass = {astro-ph.GA},
       adsurl = {https://ui.adsabs.harvard.edu/abs/2024ApJ...966L..15J}
}

@ARTICLE{2024MNRAS.528.4010O,
       author = {{O'Beirne}, T. and {Staveley-Smith}, L. and {Wong}, O.~I. and {Westmeier}, T. and {Batten}, G. and {Kilborn}, V.~A. and {Lee-Waddell}, K. and {Mancera Pi{\~n}a}, P.~E. and {Rom{\'a}n}, J. and {Verdes-Montenegro}, L. and {Catinella}, B. and {Cortese}, L. and {Deg}, N. and {D{\'e}nes}, H. and {For}, B.~Q. and {Kamphuis}, P. and {Koribalski}, B.~S. and {Murugeshan}, C. and {Rhee}, J. and {Spekkens}, K. and {Wang}, J. and {Bekki}, K. and {Lṕpez-S{\'a}nchez}, {\'A}. R.},
        title = "{WALLABY pilot survey: an 'almost' dark cloud near the Hydra cluster}",
      journal = {\mnras},
         year = 2024,
        month = mar,
       volume = {528},
       number = {3},
        pages = {4010-4028},
          doi = {10.1093/mnras/stae215},
archivePrefix = {arXiv},
       eprint = {2401.09738},
 primaryClass = {astro-ph.GA},
       adsurl = {https://ui.adsabs.harvard.edu/abs/2024MNRAS.528.4010O}
}

@ARTICLE{2025A&A...696A.185H,
       author = {{Haubner}, Konstantin and {Lelli}, Federico and {Di Teodoro}, Enrico and {Duey}, Francis and {McGaugh}, Stacy and {Schombert}, James},
        title = "{A new uncertainty scheme for galaxy distances from flow models}",
      journal = {\aap},
         year = 2025,
        month = apr,
       volume = {696},
          eid = {A185},
        pages = {A185},
          doi = {10.1051/0004-6361/202554164},
archivePrefix = {arXiv},
       eprint = {2503.08491},
 primaryClass = {astro-ph.CO},
       adsurl = {https://ui.adsabs.harvard.edu/abs/2025A&A...696A.185H}
}

@ARTICLE{2024ApJ...976...60M,
       author = {{McQuinn}, Kristen B.~W. and {Newman}, Max J.~B. and {Skillman}, Evan D. and {Telford}, O. Grace and {Brooks}, Alyson and {Adams}, Elizabeth A.~K. and {Berg}, Danielle A. and {Boyer}, Martha L. and {Cannon}, John M. and {Dolphin}, Andrew E. and {Pahl}, Anthony J. and {Rhode}, Katherine L. and {Salzer}, John J. and {Cohen}, Roger E. and {Goldman}, Steve R.},
        title = "{The Ancient Star Formation History of the Extremely Low-mass Galaxy Leo P: An Emerging Trend of a Post-reionization Pause in Star Formation}",
      journal = {\apj},
         year = 2024,
        month = nov,
       volume = {976},
       number = {1},
          eid = {60},
        pages = {60},
          doi = {10.3847/1538-4357/ad8158},
archivePrefix = {arXiv},
       eprint = {2409.19050},
 primaryClass = {astro-ph.GA},
       adsurl = {https://ui.adsabs.harvard.edu/abs/2024ApJ...976...60M}
}

@ARTICLE{2013AJ....146...15G,
       author = {{Giovanelli}, Riccardo and {Haynes}, Martha P. and {Adams}, Elizabeth A.~K. and {Cannon}, John M. and {Rhode}, Katherine L. and {Salzer}, John J. and {Skillman}, Evan D. and {Bernstein-Cooper}, Elijah Z. and {McQuinn}, Kristen B.~W.},
        title = "{ALFALFA Discovery of the Nearby Gas-rich Dwarf Galaxy Leo P. I. H I Observations}",
      journal = {\aj},
         year = 2013,
        month = jul,
       volume = {146},
       number = {1},
          eid = {15},
        pages = {15},
          doi = {10.1088/0004-6256/146/1/15},
archivePrefix = {arXiv},
       eprint = {1305.0272},
 primaryClass = {astro-ph.CO},
       adsurl = {https://ui.adsabs.harvard.edu/abs/2013AJ....146...15G}
}

@ARTICLE{2007ApJ...656L..13I,
       author = {{Irwin}, M.~J. and {Belokurov}, V. and {Evans}, N.~W. and {Ryan-Weber}, E.~V. and {de Jong}, J.~T.~A. and {Koposov}, S. and {Zucker}, D.~B. and {Hodgkin}, S.~T. and {Gilmore}, G. and {Prema}, P. and {Hebb}, L. and {Begum}, A. and {Fellhauer}, M. and {Hewett}, P.~C. and {Kennicutt}, Jr., R.~C. and {Wilkinson}, M.~I. and {Bramich}, D.~M. and {Vidrih}, S. and {Rix}, H. -W. and {Beers}, T.~C. and {Barentine}, J.~C. and {Brewington}, H. and {Harvanek}, M. and {Krzesinski}, J. and {Long}, D. and {Nitta}, A. and {Snedden}, S.~A.},
        title = "{Discovery of an Unusual Dwarf Galaxy in the Outskirts of the Milky Way}",
      journal = {\apjl},
         year = 2007,
        month = feb,
       volume = {656},
       number = {1},
        pages = {L13-L16},
          doi = {10.1086/512183},
archivePrefix = {arXiv},
       eprint = {astro-ph/0701154},
 primaryClass = {astro-ph},
       adsurl = {https://ui.adsabs.harvard.edu/abs/2007ApJ...656L..13I}
}

@ARTICLE{2007ApJ...670..313S,
       author = {{Simon}, Joshua D. and {Geha}, Marla},
        title = "{The Kinematics of the Ultra-faint Milky Way Satellites: Solving the Missing Satellite Problem}",
      journal = {\apj},
         year = 2007,
        month = nov,
       volume = {670},
       number = {1},
        pages = {313-331},
          doi = {10.1086/521816},
archivePrefix = {arXiv},
       eprint = {0706.0516},
 primaryClass = {astro-ph},
       adsurl = {https://ui.adsabs.harvard.edu/abs/2007ApJ...670..313S}
}

@ARTICLE{2008MNRAS.384..535R,
       author = {{Ryan-Weber}, Emma V. and {Begum}, Ayesha and {Oosterloo}, Tom and {Pal}, Sabyasachi and {Irwin}, Michael J. and {Belokurov}, Vasily and {Evans}, N. Wyn and {Zucker}, Daniel B.},
        title = "{The Local Group dwarf Leo T: HI on the brink of star formation}",
      journal = {\mnras},
         year = 2008,
        month = feb,
       volume = {384},
       number = {2},
        pages = {535-540},
          doi = {10.1111/j.1365-2966.2007.12734.x},
archivePrefix = {arXiv},
       eprint = {0711.2979},
 primaryClass = {astro-ph},
       adsurl = {https://ui.adsabs.harvard.edu/abs/2008MNRAS.384..535R}
}

@ARTICLE{2012ApJ...750...99T,
       author = {{Tonry}, J.~L. and {Stubbs}, C.~W. and {Lykke}, K.~R. and {Doherty}, P. and {Shivvers}, I.~S. and {Burgett}, W.~S. and {Chambers}, K.~C. and {Hodapp}, K.~W. and {Kaiser}, N. and {Kudritzki}, R. -P. and {Magnier}, E.~A. and {Morgan}, J.~S. and {Price}, P.~A. and {Wainscoat}, R.~J.},
        title = "{The Pan-STARRS1 Photometric System}",
      journal = {\apj},
         year = 2012,
        month = may,
       volume = {750},
       number = {2},
          eid = {99},
        pages = {99},
          doi = {10.1088/0004-637X/750/2/99},
archivePrefix = {arXiv},
       eprint = {1203.0297},
 primaryClass = {astro-ph.IM},
       adsurl = {https://ui.adsabs.harvard.edu/abs/2012ApJ...750...99T}
}

@ARTICLE{2006AJ....131..363D,
       author = {{de Blok}, W.~J.~G. and {Walter}, F.},
        title = "{The Star Formation Threshold in NGC 6822}",
      journal = {\aj},
         year = 2006,
        month = jan,
       volume = {131},
       number = {1},
        pages = {363-374},
          doi = {10.1086/497828},
archivePrefix = {arXiv},
       eprint = {astro-ph/0509107},
 primaryClass = {astro-ph},
       adsurl = {https://ui.adsabs.harvard.edu/abs/2006AJ....131..363D}
}

@ARTICLE{2013ApJ...765..136S,
       author = {{Stilp}, Adrienne M. and {Dalcanton}, Julianne J. and {Warren}, Steven R. and {Skillman}, Evan and {Ott}, J{\"u}rgen and {Koribalski}, B{\"a}rbel},
        title = "{Global H I Kinematics in Dwarf Galaxies}",
      journal = {\apj},
         year = 2013,
        month = mar,
       volume = {765},
       number = {2},
          eid = {136},
        pages = {136},
          doi = {10.1088/0004-637X/765/2/136},
archivePrefix = {arXiv},
       eprint = {1301.1989},
 primaryClass = {astro-ph.GA},
       adsurl = {https://ui.adsabs.harvard.edu/abs/2013ApJ...765..136S}
}

@ARTICLE{2006MNRAS.368.1479D,
       author = {{Davies}, J.~I. and {Disney}, M.~J. and {Minchin}, R.~F. and {Auld}, R. and {Smith}, R.},
        title = "{The existence and detection of optically dark galaxies by 21-cm surveys}",
      journal = {\mnras},
         year = 2006,
        month = may,
       volume = {368},
       number = {3},
        pages = {1479-1488},
          doi = {10.1111/j.1365-2966.2006.10247.x},
archivePrefix = {arXiv},
       eprint = {astro-ph/0609747},
 primaryClass = {astro-ph},
       adsurl = {https://ui.adsabs.harvard.edu/abs/2006MNRAS.368.1479D}
}

@ARTICLE{2014Natur.509..177V,
       author = {{Vogelsberger}, M. and {Genel}, S. and {Springel}, V. and {Torrey}, P. and {Sijacki}, D. and {Xu}, D. and {Snyder}, G. and {Bird}, S. and {Nelson}, D. and {Hernquist}, L.},
        title = "{Properties of galaxies reproduced by a hydrodynamic simulation}",
      journal = {\nat},
         year = 2014,
        month = may,
       volume = {509},
       number = {7499},
        pages = {177-182},
          doi = {10.1038/nature13316},
archivePrefix = {arXiv},
       eprint = {1405.1418},
 primaryClass = {astro-ph.CO},
       adsurl = {https://ui.adsabs.harvard.edu/abs/2014Natur.509..177V}
}

@ARTICLE{2015A&C....13...12N,
       author = {{Nelson}, D. and {Pillepich}, A. and {Genel}, S. and {Vogelsberger}, M. and {Springel}, V. and {Torrey}, P. and {Rodriguez-Gomez}, V. and {Sijacki}, D. and {Snyder}, G.~F. and {Griffen}, B. and {Marinacci}, F. and {Blecha}, L. and {Sales}, L. and {Xu}, D. and {Hernquist}, L.},
        title = "{The illustris simulation: Public data release}",
      journal = {Astronomy and Computing},
         year = 2015,
        month = nov,
       volume = {13},
        pages = {12-37},
          doi = {10.1016/j.ascom.2015.09.003},
archivePrefix = {arXiv},
       eprint = {1504.00362},
 primaryClass = {astro-ph.CO},
       adsurl = {https://ui.adsabs.harvard.edu/abs/2015A&C....13...12N}
}

@ARTICLE{2005MNRAS.363L..21B,
       author = {{Bekki}, Kenji and {Koribalski}, B{\"a}rbel S. and {Kilborn}, Virginia A.},
        title = "{Dark galaxies or tidal debris? Kinematical clues to the origin of massive isolated HI clouds}",
      journal = {\mnras},
         year = 2005,
        month = oct,
       volume = {363},
       number = {1},
        pages = {L21-L25},
          doi = {10.1111/j.1745-3933.2005.00076.x},
archivePrefix = {arXiv},
       eprint = {astro-ph/0505580},
 primaryClass = {astro-ph},
       adsurl = {https://ui.adsabs.harvard.edu/abs/2005MNRAS.363L..21B}
}

@ARTICLE{2017MNRAS.467.3648T,
       author = {{Taylor}, R. and {Davies}, J.~I. and {J{\'a}chym}, P. and {Keenan}, O. and {Minchin}, R.~F. and {Palou{\v{s}}}, J. and {Smith}, R. and {W{\"u}nsch}, R.},
        title = "{Kinematic clues to the origins of starless H I clouds: dark galaxies or tidal debris?}",
      journal = {\mnras},
         year = 2017,
        month = may,
       volume = {467},
       number = {3},
        pages = {3648-3661},
          doi = {10.1093/mnras/stx187},
archivePrefix = {arXiv},
       eprint = {1701.05361},
 primaryClass = {astro-ph.GA},
       adsurl = {https://ui.adsabs.harvard.edu/abs/2017MNRAS.467.3648T}
}

@ARTICLE{2022A&A...658A.146V,
       author = {{van Cappellen}, W.~A. and {Oosterloo}, T.~A. and {Verheijen}, M.~A.~W. and {Adams}, E.~A.~K. and {Adebahr}, B. and {Braun}, R. and {Hess}, K.~M. and {Holties}, H. and {van der Hulst}, J.~M. and {Hut}, B. and {Kooistra}, E. and {van Leeuwen}, J. and {Loose}, G.~M. and {Morganti}, R. and {Moss}, V.~A. and {Orr{\'u}}, E. and {Ruiter}, M. and {Schoenmakers}, A.~P. and {Vermaas}, N.~J. and {Wijnholds}, S.~J. and {van Amesfoort}, A.~S. and {Arts}, M.~J. and {Attema}, J.~J. and {Bakker}, L. and {Bassa}, C.~G. and {Bast}, J.~E. and {Benthem}, P. and {Beukema}, R. and {Blaauw}, R. and {de Blok}, W.~J.~G. and {Bouwhuis}, M. and {van den Brink}, R.~H. and {Connor}, L. and {Coolen}, A.~H.~W.~M. and {Damstra}, S. and {van Diepen}, G.~N.~J. and {de Goei}, R. and {D{\'e}nes}, H. and {Drost}, M. and {Ebbendorf}, N. and {Frank}, B.~S. and {Gardenier}, D.~W. and {Gerbers}, M. and {Grange}, Y.~G. and {Grit}, T. and {Gunst}, A.~W. and {Gupta}, N. and {Ivashina}, M.~V. and {J{\'o}zsa}, G.~I.~G. and {Janssen}, G.~H. and {Koster}, A. and {Kruithof}, G.~H. and {Kuindersma}, S.~J. and {Kutkin}, A. and {Lucero}, D.~M. and {Maan}, Y. and {Maccagni}, F.~M. and {van der Marel}, J. and {Mika}, A. and {Morawietz}, J. and {Mulder}, H. and {Mulder}, E. and {Norden}, M.~J. and {Offringa}, A.~R. and {Oostrum}, L.~C. and {Overeem}, R.~E. and {Paragi}, Z. and {Pepping}, H.~J. and {Petroff}, E. and {Pisano}, D.~J. and {Polatidis}, A.~G. and {Prasad}, P. and {de Reijer}, J.~P.~R. and {Romein}, J.~W. and {Schaap}, J. and {Schoonderbeek}, G.~W. and {Schulz}, R. and {van der Schuur}, D. and {Sclocco}, A. and {Sluman}, J.~J. and {Smits}, R. and {Stappers}, B.~W. and {Straal}, S.~M. and {Stuurwold}, K.~J.~C. and {Verstappen}, J. and {Vohl}, D. and {Wierenga}, K.~J. and {Woestenburg}, E.~E.~M. and {Zanting}, A.~W. and {Ziemke}, J.},
        title = "{Apertif: Phased array feeds for the Westerbork Synthesis Radio Telescope. System overview and performance characteristics}",
      journal = {\aap},
         year = 2022,
        month = feb,
       volume = {658},
          eid = {A146},
        pages = {A146},
          doi = {10.1051/0004-6361/202141739},
archivePrefix = {arXiv},
       eprint = {2109.14234},
 primaryClass = {astro-ph.IM},
       adsurl = {https://ui.adsabs.harvard.edu/abs/2022A&A...658A.146V}
}

@ARTICLE{2022A&A...667A..38A,
       author = {{Adams}, E.~A.~K. and {Adebahr}, B. and {de Blok}, W.~J.~G. and {D{\'e}nes}, H. and {Hess}, K.~M. and {van der Hulst}, J.~M. and {Kutkin}, A. and {Lucero}, D.~M. and {Morganti}, R. and {Moss}, V.~A. and {Oosterloo}, T.~A. and {Orr{\'u}}, E. and {Schulz}, R. and {van Amesfoort}, A.~S. and {Berger}, A. and {Boersma}, O.~M. and {Bouwhuis}, M. and {van den Brink}, R. and {van Cappellen}, W.~A. and {Connor}, L. and {Coolen}, A.~H.~W.~M. and {Damstra}, S. and {van Diepen}, G.~N.~J. and {Dijkema}, T.~J. and {Ebbendorf}, N. and {Grange}, Y.~G. and {de Goei}, R. and {Gunst}, A.~W. and {Holties}, H.~A. and {Hut}, B. and {Ivashina}, M.~V. and {J{\'o}zsa}, G.~I.~G. and {van Leeuwen}, J. and {Loose}, G.~M. and {Maan}, Y. and {Mancini}, M. and {Mika}, {\'A}. and {Mulder}, H. and {Norden}, M.~J. and {Offringa}, A.~R. and {Oostrum}, L.~C. and {Pastor-Marazuela}, I. and {Pisano}, D.~J. and {Ponomareva}, A.~A. and {Romein}, J.~W. and {Ruiter}, M. and {Schoenmakers}, A.~P. and {van der Schuur}, D. and {Sluman}, J.~J. and {Smits}, R. and {Stuurwold}, K.~J.~C. and {Verstappen}, J. and {Vilchez}, N.~P.~E. and {Vohl}, D. and {Wierenga}, K.~J. and {Wijnholds}, S.~J. and {Woestenburg}, E.~E.~M. and {Zanting}, A.~W. and {Ziemke}, J.},
        title = "{First release of Apertif imaging survey data}",
      journal = {\aap},
         year = 2022,
        month = nov,
       volume = {667},
          eid = {A38},
        pages = {A38},
          doi = {10.1051/0004-6361/202244007},
archivePrefix = {arXiv},
       eprint = {2208.05348},
 primaryClass = {astro-ph.IM},
       adsurl = {https://ui.adsabs.harvard.edu/abs/2022A&A...667A..38A}
}

@ARTICLE{2002ApJS..142..161P,
       author = {{Pisano}, D.~J. and {Wilcots}, Eric M. and {Liu}, Charles T.},
        title = "{An H I/Optical Atlas of Isolated Galaxies}",
      journal = {\apjs},
         year = 2002,
        month = oct,
       volume = {142},
       number = {2},
        pages = {161-222},
          doi = {10.1086/341787},
       adsurl = {https://ui.adsabs.harvard.edu/abs/2002ApJS..142..161P}
}

\newpage

\begin{appendix}

\section{Channel maps}
\label{c3:app:chan_maps}

\begin{figure*}
    \centering
    \includegraphics[width=1\linewidth]{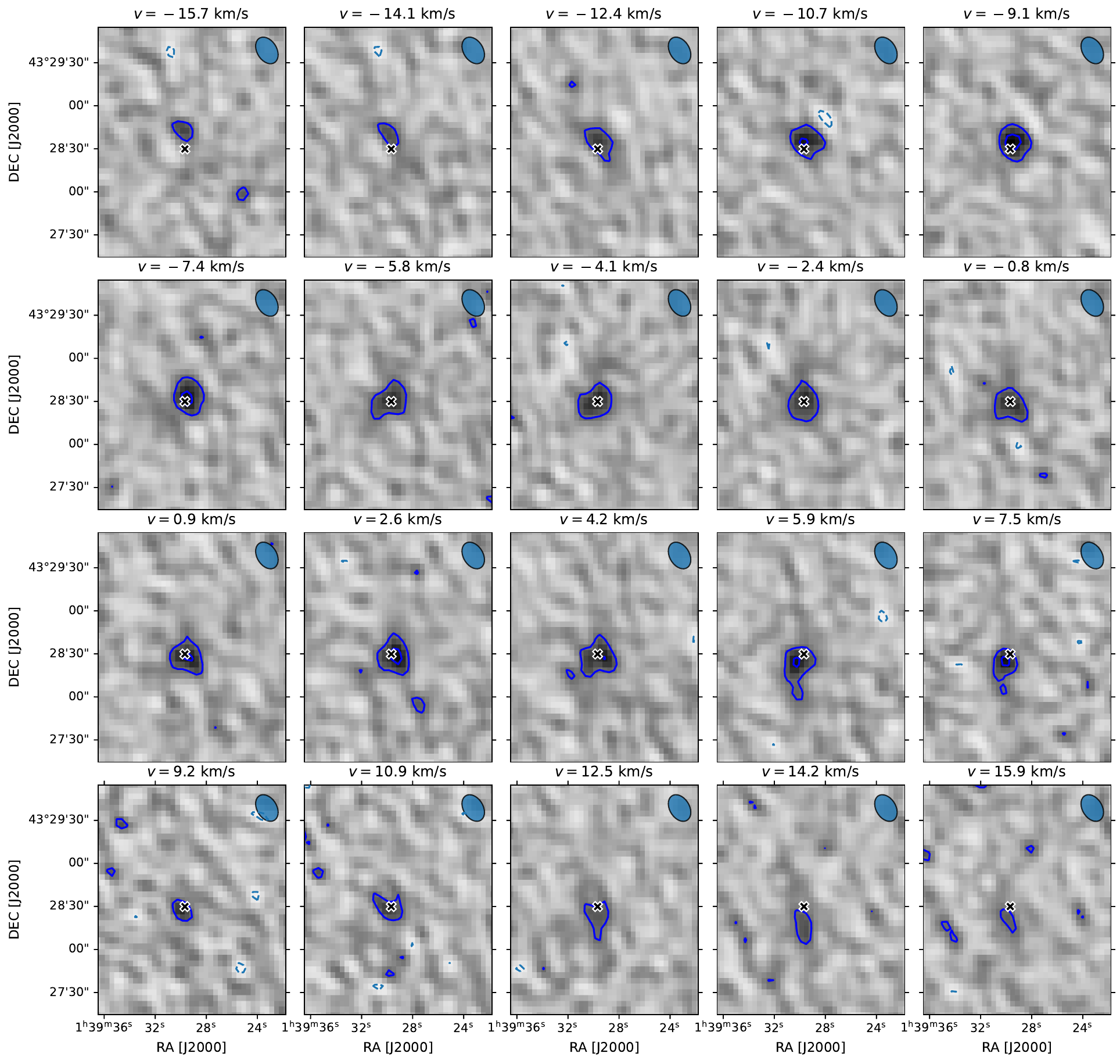}
    \caption{Channel maps of the VLA detection in the rest frame of the source. Blue contours represent the data, with dark blue lines denoting positive emission and light dashed blue lines the negative. Contours are plotted starting from 3 times the noise in the cube (0.65 mJy beam$^{-1}$), and are spaced by a factor of 2 in intensity. The black $X$ indicates the \hi\ centroid of the galaxy.}
    \label{p3:fig:channel_maps}
\end{figure*}

In Fig. \ref{p3:fig:channel_maps}, we show the channel maps of the VLA detection containing emission from the galaxy.

\end{appendix}


\end{document}